\documentclass[acmsmall,screen]{acmart}
\usepackage[ruled,lined,linesnumbered,vlined]{algorithm2e}

\SetCommentSty{mycommfont}
\DontPrintSemicolon %

\usepackage{amsmath}
\usepackage{amsfonts}
\usepackage{amsthm}
\usepackage{balance}
\usepackage{enumitem}
\usepackage{graphicx}
\usepackage{listings}
\usepackage{multicol}
\usepackage{multirow}
\usepackage{natbib}
\usepackage{subcaption}
\usepackage{url}
\usepackage{xspace}
\usepackage{xcolor}
\usepackage{tcolorbox}
\usepackage{colortbl}

\usepackage{tikz}
\usepackage{pgfplots}
\newcommand*\circled[1]{\tikz[baseline=(char.base)]{
    \node[shape=circle,draw,inner sep=0.5pt] (char) {\small#1};}}
\usetikzlibrary{shapes}
\usetikzlibrary{shapes.geometric}
\usetikzlibrary{arrows.meta, positioning}

\newcommand{\eg}{\mbox{e.g.}\xspace}
\newcommand{\etal}{\mbox{et al.}\xspace}

\newcommand{\ie}{\mbox{i.e.}\xspace}

\definecolor{BlindColorTolOne}{HTML}{332288}
\definecolor{BlindColorTolTwo}{HTML}{117733} %
\definecolor{BlindColorTolThree}{HTML}{44AA99}
\definecolor{BlindColorTolFour}{HTML}{88CCEE}
\definecolor{BlindColorTolFive}{HTML}{DDCC77}
\definecolor{BlindColorTolSix}{HTML}{CC6677} %
\definecolor{BlindColorTolSeven}{HTML}{AA4499}
\definecolor{BlindColorTolEight}{HTML}{882255}

\definecolor{BlindColorWongOne}{HTML}{000000} %
\definecolor{BlindColorWongTwo}{HTML}{E69F00}
\definecolor{BlindColorWongThree}{HTML}{56B4E9}
\definecolor{BlindColorWongFour}{HTML}{009E73}
\definecolor{BlindColorWongFive}{HTML}{F0E442}
\definecolor{BlindColorWongSix}{HTML}{0072B2} %
\definecolor{BlindColorWongSeven}{HTML}{D55E00}
\definecolor{BlindColorWongEight}{HTML}{CC79A7}

\definecolor{mygreen}{HTML}{02818a}

\mathchardef\mhyphen="2D

\newcounter{FindingCounter}

\newcommand{\mycode}[1]{\texttt{#1}\xspace}

\newcommand{\projectname}[1]{\textsf{\mbox{#1}}\xspace}
\SetKwData{AlgTrue}{true}
\SetKwData{AlgFalse}{false}

\definecolor{BlindColorTolNine}{HTML}{F76724}

\newcommand{\myFinding}[1]{
    \begin{tcolorbox}[
        left=1.4mm,             %
        right=1.4mm,            %
        top=1mm,              %
        bottom=1mm,           %
        boxsep=1.2mm,           %
        arc=1.2mm,              %
        auto outer arc=false
    ]
        \textbf{RQ\refstepcounter{FindingCounter}\theFindingCounter}: #1
    \end{tcolorbox}
}

\newcommand{\myTightParagraph}[1]{
  \vspace*{0.04cm}
  \noindent \textit{\textbf{#1.}}\quad
}

\newcommand{\csmith}{CSmith\xspace}

\newcommand{\gcc}{GCC\xspace}
\newcommand{\llvm}{LLVM\xspace}

\newcommand{\legofuzz}{LegoFuzz\xspace}

\newcommand{\tamer}{\textsf{Tamer}\xspace}
\newcommand{\transformer}{\textsf{Transformer}\xspace}
\newcommand{\dthree}{\textsf{D3}\xspace}
\newcommand{\bisectionSole}{\textsf{Bisection-Sole}\xspace}

\newcommand{\proj}{\projectname{BugLens}}

\newcommand{\benchmarkGCCFourThreeZeroSize}{1,235\xspace}
\newcommand{\benchmarkGCCFourThreeZeroBugs}{29\xspace}
\newcommand{\benchmarkGCCFourFourZeroSize}{647\xspace}
\newcommand{\benchmarkGCCFourFourZeroBugs}{11\xspace}
\newcommand{\benchmarkGCCFourFiveZeroSize}{26\xspace}
\newcommand{\benchmarkGCCFourFiveZeroBugs}{7\xspace}
\newcommand{\benchmarkLLVMTwoEightZeroSize}{80\xspace}
\newcommand{\benchmarkLLVMTwoEightZeroBugs}{5\xspace}

\newcommand{\benchmarkGCCThirdteenSize}{42\xspace}
\newcommand{\benchmarkGCCThirdteenBugs}{7\xspace}

\newcommand{\testInput}{\ensuremath{\textit{P}}\xspace}

\newcommand{\testInputAlpha}{\ensuremath{\textit{P}^{\,\alpha}}\xspace}
\newcommand{\testInputBeta}{\ensuremath{\textit{P}^{\,\beta}}\xspace}
\newcommand{\goodVersion}{\ensuremath{\textit{V}_{\checkmark}}\xspace}
\newcommand{\badVersion}{\ensuremath{\textit{V}_{\times}}\xspace}

\newcommand{\commit}{\ensuremath{\textit{c}}\xspace}
\newcommand{\rankedList}{\ensuremath{\textit{L}}\xspace}

\newcommand{\optLevelOfInputI}{\ensuremath{\mathcal{O}}\xspace}

\newcommand{\ddmin}{\ensuremath{\textsf{ddmin}}\xspace}

\newcommand{\commitAlpha}{\ensuremath{\textit{c}^{\,\alpha}}\xspace}
\newcommand{\commitBeta}{\ensuremath{\textit{c}^{\,\beta}}\xspace}
\newcommand{\vectorAlpha}{\ensuremath{\mathbf{v}^{\alpha}}\xspace}
\newcommand{\vectorBeta}{\ensuremath{\mathbf{v}^{\beta}}\xspace}

\newcommand{\bisectionSoleFindAllBugs}{1027.17\xspace}
\newcommand{\bisectionWithOptFindAllBugs}{176.23\xspace}

\newcommand{\bisectionWithOptFindAllBugsOrig}{658.38\xspace}
\newcommand{\bisectionWithOptSaveEffortThanRandomOrig}{41.33\%\xspace}

\newcommand{\tamerFindAllBugs}{489.83\xspace}
\newcommand{\dthreeFindAllBugs}{453.21\xspace}

\newcommand{\bisectionOptFindAllBugsSaveEffortThanTamer}{64.02\%\xspace}
\newcommand{\bisectionOptFindAllBugsSaveEffortThanDthree}{61.12\%\xspace}

\newcommand{\numAllBuugsInTable}{54\xspace}
\newcommand{\numBugsBetterThanTamer}{48\xspace}
\newcommand{\saveEffortAvgTamer}{20.98\xspace}
\newcommand{\saveEffortAvgPercentageTamer}{33.56\%\xspace}
\newcommand{\tamerBetterCasesSaveEffortAvg}{0.26\xspace}
\newcommand{\numBugsBetterThanDthree}{40\xspace}
\newcommand{\saveEffortAvgDthree}{6.75\xspace}
\newcommand{\saveEffortAvgPercentageDthree}{10.68\%\xspace}
\newcommand{\dthreeBetterCasesSaveEffortAvg}{1.53\xspace}

\newcommand{\tamerPValueGccFourThreeZero}{0.0001\xspace}
\newcommand{\dthreePValueGccFourThreeZero}{0.03\xspace}
\newcommand{\tamerPValueGccFourFourZero}{0.002\xspace}

\newcommand{\rqOne}{How effective is using bisection as the sole criterion to deduplicate compiler bugs?}
\newcommand{\rqTwo}{How can false negatives be mitigated when leveraging bisection for compiler bug deduplication?}
\newcommand{\rqThree}{Can a bisection-based bug deduplication approach effectively function without test input minimization?}
\newcommand{\rqFour}{How efficient is the bisection-based bug deduplication in practical settings?}

\usepackage{cleveref} %

\Crefname{algocf}{Algorithm}{Algorithms}
\crefname{algocf}{Algorithm}{Algorithms}

\Crefname{algorithm}{Algorithm}{Algorithms}
\crefname{algorithm}{Algorithm}{Algorithms}

\crefname{appendix}{Appendix}{Appendices}
\Crefname{appendix}{Appendix}{Appendices}

\Crefname{figure}{Figure}{Figures}
\crefname{figure}{Figure}{Figures}

\crefname{listing}{Listing}{Listings}
\Crefname{listing}{Listing}{Listings}

\Crefname{table}{Table}{Tables}
\crefname{table}{Table}{Tables}

\crefname{thm}{Theorem}{Theorems}
\Crefname{thm}{Theorem}{Theorems}

\crefname{equation}{Equation}{Equations}
\Crefname{equation}{Equation}{Equations}

\crefformat{chapter}{\S~#2#1#3}
\crefmultiformat{chapter}{\S\S~#2#1#3}{ and~#2#1#3}{, #2#1#3}{, and~#2#1#3}

\crefformat{section}{\S~#2#1#3}
\crefmultiformat{section}{\S\S~#2#1#3}{ and~#2#1#3}{, #2#1#3}{, and~#2#1#3}

\AtBeginDocument{%
  \providecommand\BibTeX{{%
    \normalfont B\kern-0.5em{\scshape i\kern-0.25em b}\kern-0.8em\TeX}}}

\setcopyright{acmlicensed}
\copyrightyear{2018}
\acmYear{2018}
\acmDOI{XXXXXXX.XXXXXXX}

\acmConference[Conference acronym 'XX]{Make sure to enter the correct
    conference title from your rights confirmation emai}{June 03--05,
    2018}{Woodstock, NY}
\acmISBN{978-1-4503-XXXX-X/18/06}

\begin{document}

\title{On the Feasibility of Deduplicating Compiler Bugs with Bisection}

\author{Xintong Zhou}
\orcid{0009-0002-6444-5431}
\email{x27zhou@uwaterloo.ca}
\affiliation{
    \institution{University of Waterloo}
    \streetaddress{200 University Ave W}
    \city{Waterloo}
    \state{ON}
    \country{Canada}
    \postcode{N2L 3G1}
}

\author{Zhenyang Xu}
\orcid{0000-0002-9451-4031}
\email{zhenyang.xu@uwaterloo.ca}
\affiliation{
    \institution{University of Waterloo}
    \streetaddress{200 University Ave W}
    \city{Waterloo}
    \state{ON}
    \country{Canada}
    \postcode{N2L 3G1}
}

\author{Yongqiang Tian}
\orcid{0000-0003-1644-2965}
\affiliation{%
	\institution{Monash University}
	\city{Melbourne}
	\country{Australia}}
\email{yongqiang.tian@monash.edu}

\author{Chengnian Sun}
\orcid{0000-0002-0862-2491}
\email{cnsun@uwaterloo.ca}
\affiliation{
    \institution{University of Waterloo}
    \streetaddress{200 University Ave W}
    \city{Waterloo}
    \state{ON}
    \country{Canada}
    \postcode{N2L 3G1}
}

\begin{abstract}
  Random testing has proven to be an effective technique for compiler validation.
However, the debugging of bugs identified through random testing
presents a significant challenge due to the frequent occurrence of
duplicate test programs that expose identical compiler bugs.
The process to identify duplicates is a practical research problem
known as bug deduplication. Prior methodologies for compiler
bug deduplication primarily rely on program analysis to
extract bug-related features for duplicate identification,
which can result in substantial
computational overhead and limited generalizability.

This paper investigates the feasibility of employing bisection,
a standard debugging procedure largely overlooked in prior research
on compiler bug deduplication, for this purpose.
Our study demonstrates that the utilization of bisection to locate
failure-inducing commits provides a valuable criterion for deduplication,
albeit one that requires supplementary techniques for more accurate
identification.
Building on these results,
we introduce \proj,
a novel deduplication method that primarily uses bisection,
enhanced by the identification of bug-triggering optimizations to minimize false negatives.
Empirical evaluations conducted on five real-world datasets
demonstrate that \proj significantly outperforms the
state-of-the-art analysis-based methodologies \tamer and \dthree
by saving an average of \saveEffortAvgPercentageTamer and
\saveEffortAvgPercentageDthree human effort to identify the
same number of distinct bugs.
Given the inherent simplicity and generalizability of bisection,
it presents a highly practical solution for compiler bug deduplication
in real-world applications.

\end{abstract}

\begin{CCSXML}
<ccs2012>
 <concept>
  <concept_id>10010520.10010553.10010562</concept_id>
  <concept_desc>Computer systems organization~Embedded systems</concept_desc>
  <concept_significance>500</concept_significance>
 </concept>
 <concept>
  <concept_id>10010520.10010575.10010755</concept_id>
  <concept_desc>Computer systems organization~Redundancy</concept_desc>
  <concept_significance>300</concept_significance>
 </concept>
 <concept>
  <concept_id>10010520.10010553.10010554</concept_id>
  <concept_desc>Computer systems organization~Robotics</concept_desc>
  <concept_significance>100</concept_significance>
 </concept>
 <concept>
  <concept_id>10003033.10003083.10003095</concept_id>
  <concept_desc>Networks~Network reliability</concept_desc>
  <concept_significance>100</concept_significance>
 </concept>
</ccs2012>
\end{CCSXML}

\ccsdesc[500]{Computer systems organization~Embedded systems}
\ccsdesc[300]{Computer systems organization~Redundancy}
\ccsdesc{Computer systems organization~Robotics}
\ccsdesc[100]{Networks~Network reliability}

\keywords{compiler bug deduplication, bisection, random testing, debugging}

\setcopyright{none} %
\settopmatter{printacmref=false} %
\renewcommand\footnotetextcopyrightpermission[1]{}

\maketitle

\section{Introduction}
\label{sec:intro}
Compilers are among the most fundamental and critical software
systems, underpinning nearly all modern software.
Hence ensuring compiler correctness is crucial to guarantee
the reliability of software built upon them.
However, the increasing complexity of modern compilers
makes them susceptible to various bugs.
These bugs can result in incorrect code generation,
leading to unexpected program behavior and potential security vulnerabilities.
Therefore, detecting and resolving compiler bugs is paramount
to maintaining the reliability and correctness of software systems.

Random testing has been proven highly effective for
improving compiler correctness by discovering numerous bugs~\cite{csmith,emi,sun2016finding,yarpgen,yarpgen2.0,grayc,creal}.
For instance, CSmith~\cite{csmith} successfully discovered
hundreds of bugs in GCC and LLVM by randomly generating
C programs to perform differential testing.
By producing numerous test programs, these techniques
effectively uncover compiler bugs.
However, in practice, a single bug may manifest through various
test programs, resulting in redundant
programs that trigger the same
underlying bug.
Given the substantial volume of test programs, classifying these programs
based on the specific bugs they expose, referred to as \emph{bug deduplication},
remains challenging yet crucial for prioritizing bug reports and
reducing developers' debugging effort.
In particular, unlike crash bugs, which can be readily deduplicated
using crash information~\cite{tamer,d3},
\emph{miscompilation bugs} (also known as
wrong-code bugs) lack such distinguishing characteristics,
making deduplication significantly more difficult~\cite{tamer,d3}.

In the literature, several techniques have been proposed to address
the bug deduplication problem in compilers~\cite{tamer,transformer,d3}.
These techniques are typically \emph{analysis-based},
relying on collecting program features and runtime information
to measure similarity among different test cases.
For example, Chen \etal~\cite{tamer} proposed \tamer,
which distinguishes bug-triggering programs by analyzing
compiler code coverage during test execution.
Subsequently, Yang \etal introduced \dthree~\cite{d3},
integrating more detailed static and runtime information
to improve bug deduplication accuracy.
While these approaches demonstrate effectiveness in certain contexts,
they often involve significant implementation complexity and
dependencies on external tools, substantially limiting their practical
applicability and generality.
In particular, these techniques always depend on
test input minimization tools~\cite{creduce,perses,deltadebugging}
to clear irrelevant
code and reduce the analysis complexity,
which introduces additional overhead to the deduplication process.
Thus, a more \emph{lightweight, generalizable, and independent} approach
for bug deduplication is highly desirable in real-world scenarios.

Modern software projects (\eg, \gcc and \llvm) extensively employ
version control systems like Git~\cite{git} to manage development activities.
These version histories document all the code changes,
thus naturally including the introduction of bugs.
In practice, developers frequently perform binary search, \ie, bisection,
on the commit history to identify the exact commit that causes the failure,
facilitating efficient root-cause analysis and bug resolution.
While bisection is widely recognized as a standard debugging technique
for its simplicity and effectiveness, leveraging it explicitly for
bug deduplication has remained largely overlooked and
unexplored in the literature.

To bridge this gap, we conduct the first systematic study on the
feasibility of employing bisection for deduplicating compiler bugs,
specifically focusing on miscompilation bugs.
The fundamental insight guiding our study is that test failures
triggered by separate code changes typically indicate different
root causes, \ie, distinct bugs.
Thus, the commit identified through bisection, \ie, the first commit
exhibiting the bug, naturally serves as a criterion to
deduplicate various bug-triggering test programs.
However, given multiple bugs can be introduced or triggered by a
single commit---a common scenario in compiler development---relying
solely on this criterion raises concerns about that distinct bugs
may be mistakenly identified as identical
(referred to as false negatives in \cref{subsec:rq2}).
Determining whether such false negatives can be effectively minimized
without substantial additional effort is essential for establishing
the practicality of bisection-based deduplication.
Furthermore, the generality and efficiency of this approach are
critical factors determining its applicability in real-world scenarios.
To systematically evaluate these aspects,
we formulate the following research questions (RQs):
\begin{itemize}[leftmargin=*,topsep=5pt]
\item RQ1: \rqOne
\item RQ2: \rqTwo
\item RQ3: \rqThree
\item RQ4: \rqFour
\end{itemize}

RQ1 and RQ2 focus on effectiveness, RQ3 evaluates generality by analyzing
dependence on
test input minimization, and RQ4 assesses practical efficiency.
By answering these research questions,
we aim to present a comprehensive evaluation on the feasibility
of leveraging bisection to deduplicate compiler bugs.
To assist the practical application,
based on our study findings,
we propose \proj, a bisection-based bug deduplication
approach based on our study findings.
\proj leverages the failure-inducing commit identified through bisection
as the primary criterion, accompanied with examining the bug-triggering
optimizations as a subsidiary factor to mitigate false negatives.
Our empirical evaluations demonstrate that \proj
significantly outperforms two state-of-the-art analysis-based
techniques, \tamer~\cite{tamer} and \dthree~\cite{d3},
by saving \saveEffortAvgPercentageTamer and
\saveEffortAvgPercentageDthree of the human effort on average to
identify the same number of unique bugs, respectively.
Additionally, results from RQ3 confirm that the approach remains largely
effective with unminimized test programs, thus providing a valuable
alternative when effective test input minimization is unavailable.
RQ4 confirms the practical efficiency of our approach.

Our study demonstrates that the bisection-based approach is a promising
alternative to existing analysis-based techniques for compiler bug deduplication,
releasing the task from the excessive complexity and limited generality.
Our study emphasizes the importance of re-evaluating simple
techniques before resorting to more elaborate solutions.
Overall, this paper makes the following contributions:
\begin{itemize}[leftmargin=*]

    \item We conduct the first systematic study on the practical feasibility, including
    effectiveness, generality, and efficiency, of employing bisection for compiler bug deduplication.

    \item Based on our study findings, we propose \proj, a bisection-based bug deduplication
    approach, which combines the bisection results with the bug-triggering
    optimizations to provide effective bug deduplication.

    \item Empirical evaluations conducted on five real-world datasets demonstrate
    that \proj significantly outperforms the state-of-the-art analysis-based
    methodologies in deduplication effectiveness, while exhibiting greater simplicity and generality.

    \item To promise the reproducibility of our study and facilitate future research,
    we have made the artifact publicly available
    at
    \href{https://anonymous.4open.science/r/BugLens-Artifact/}{\texttt{https://anonymous.4open.science/r/BugLens-Artifact/}}

\end{itemize}

\section{Preliminaries}
\label{sec:background}
This section provides essential background information required to
understand compiler bug deduplication and underscores
the limitations of existing approaches.

\subsection{Challenges in Random Testing}
\label{subsec:bug-deduplication}
Random testing enhances compiler correctness by automatically
generating large amounts of  random test programs to expose
compiler bugs~\cite{csmith,emi,sun2016finding,yarpgen}.
Although random testing has proven effective at identifying new bugs,
it faces two major challenges in practice.

First, randomly generated test programs that trigger compiler bugs
are often large and complex, typically containing a substantial
amount of bug-irrelevant code, which complicates the debugging process.
A recent study~\cite{sun2016toward} on bug characteristics
in GCC and LLVM reports that minimized test programs
average only about 30 lines of code to trigger compiler bugs,
whereas the original, non-minimized programs frequently span hundreds
or even thousands of lines. Automated test input minimization techniques,
such as Delta Debugging~\cite{deltadebugging}, have proven effective
in alleviating this issue. However, the process remains computationally
expensive, often requiring hours to minimize a single program~\cite{creduce,perses}.

The second issue is that program generators tend to
produce numerous duplicate programs which, despite syntactic differences,
trigger the same underlying bug.
For instance, in one dataset used in our study,
\benchmarkGCCFourThreeZeroSize C programs generated by
\csmith~\cite{csmith}, a state-of-the-art C program generator,
trigger only \benchmarkGCCFourThreeZeroBugs uniques bugs
in the \gcc-4.3.0 compiler.
Deduplicating these programs and selecting proper ones for
investigation are crucial for efficient debugging.
The objective is to identify as many distinct bugs as possible while
minimizing the number of test programs developers need to analyze.
Ideally, with an effective deduplication strategy,
only \benchmarkGCCFourThreeZeroBugs programs would be required to
manifest all of these bugs.
Effective bug deduplication would significantly reduce the
number of cases developers need to investigate,
saving considerable time and effort.

\subsection{Analysis-Based Compiler Bug Deduplication}
\label{subsec:existing-techniques}
In the literature, several techniques have been proposed to tackle
the bug deduplication problem in compilers~\cite{tamer,transformer,d3}.
A common strategy among these techniques is to reformulate
bug deduplication as a test case prioritization problem.
Specifically, by assessing the similarity among test cases,
these approaches prioritize cases likely to reveal distinct bugs.
When measuring the similarity between test cases, these
approaches are typically \emph{analysis-based}, collecting
various program features and runtime information
to predict if two test cases trigger the same underlying bug.
The state-of-the-art techniques include:

\myTightParagraph{\tamer}
Chen \etal~\cite{tamer} are the first to systematically formulate
and study the bug deduplication problem in compiler testing.
They proposed a technique named \tamer,
which measures the distance between
pairs of bug-triggering programs using various static program features
(\eg, types, operators and loops) and runtime information
(\eg, execution output, line and function coverage).
Based on the distance, \tamer prioritizes test programs that are more
distant from others, as they are more likely to trigger distinct bugs.
Their evaluation showed that \tamer achieved optimal performance with
using \emph{compiler function coverage} as the distance metric.

\myTightParagraph{\transformer}
Donaldson \etal~\cite{transformer} proposed a technique,
referred to as \transformer,
specifically tailored for transformation-based compiler testing.
The fundamental insight of \transformer is that two bug-triggering programs
generated by identical transformation operations tend to trigger the same bug.
While \transformer incurs minimal additional cost — since transformation
operations are inherently part of the testing process — its practical
utility is limited due to its narrow focus on transformation-based testing.
Another study~\cite{d3} further demonstrated that \transformer performs
poorly when extended to general compiler testing scenarios.

\myTightParagraph{\dthree}
Most recently, Yang \etal~\cite{d3} introduce \dthree, which
performs a three-dimensional analysis to enhance the effectiveness of bug
deduplication. Their approach is grounded on the observation that three
distinct dimensions of information are crucial for effectively
characterizing compiler bugs:
\circled{1} the program features of the test program triggering the bug,
\circled{2} the compiler optimizations necessary to reproduce the bug, and
\circled{3} the execution information of the compiler on the program.
\dthree assesses the similarity between test programs using a
three-dimensional distance metric that integrates these dimensions.
Although collecting detailed information facilitates more effective deduplication,
it also increases the complexity and overhead of the process.

While these techniques are theoretically sound and show
acceptable results in certain scenarios, inherent limitations
restrict their practical applicability in real-world scenarios.

\myTightParagraph{Complex Implementation and Limited Generality}
Analysis-based methods rely heavily on collecting detailed program
and execution features to measure the similarity between test cases.
Such implementations are typically complex and heavily
dependent on external tools.
For example, \tamer and \dthree both utilize Gcov~\cite{gcov} to collect
coverage information,
and \dthree additionally requires tree-sitter~\cite{treesitter}
for AST transformations, GumTree~\cite{gumTree} for AST difference
extraction, and spectrum-based fault localization (SBFL)
tools~\cite{sohn2017fluccs} for identifying buggy code.
These dependencies, often language-specific,
limit the generality of these methods,
requiring substantial effort to adapt them to different domains.
Although \transformer is less dependent on detailed program analysis,
its scope is narrowly confined to transformation-based testing scenarios.
\emph{In this context, a more generalizable bug deduplication approach
applicable across diverse domains is highly desirable.}

\myTightParagraph{Strong Reliance on Test Input Minimization}
As discussed in \cref{subsec:bug-deduplication},
compiler fuzzers typically generate large and complex programs
filled with bug-irrelevant code, which complicate the debugging process.
These extraneous elements also hinder the effectiveness
of current analysis-based bug deduplication methods,
making them unable to provide prominent improvements over
a random selection baseline when handling
unminimized programs~\cite{tamer,d3}.
In particular, Chen \etal~\cite{tamer}
explicitly emphasize the necessity of test input minimization for
effective bug deduplication:
\begin{quote}
    \emph{"Our view is that attempting to tame a fuzzer without the aid of a
    solid test-case reducer is inadvisable."}

    \quad\quad\quad\quad\quad\quad\quad\
    \quad\quad\quad\quad\quad\quad\quad\
    \quad\quad\quad\quad\quad\quad
    --- Chen \etal in \tamer's paper~\cite{tamer}
\end{quote}
Moreover, \dthree is even infeasible to apply to unminimized test programs.
For instance, a critical step in \dthree is to mutate a bug-triggering program
into a passing one, which is nearly impossible to success on large and complex programs.
As a result, both \tamer and \dthree require test input minimization
as a prerequisite, relying on external tools such as
C-Reduce~\cite{creduce} and Perses~\cite{perses}.

However, test input minimization is time-consuming and computationally
expensive. For example, it could take hours to minimize a single
bug-triggering C program generated by \csmith~\cite{csmith},
with state-of-the-art reducers like C-Reduce~\cite{creduce}.
To this end, minimizing every test input for the purpose of
deduplication becomes undesirable.
Ideally, deduplication would occur independently of minimization,
with only a small subset of test inputs needed to be
minimized for reporting and debugging purposes.
Given the high volume of duplicate test inputs,
minimization has become a critical bottleneck in current
analysis-based deduplication approaches.
Additionally, bug slippage~\cite{tamer}---where the original
and minimized versions of a test case trigger different bugs---may
occur during minimization, further complicating the process.
\emph{In this context, a deduplication approach less reliant on
test input minimization would be significantly beneficial.}

\subsection{Version Control Systems and Bisection}
Version Control Systems (VCSs)~\cite{spinellis2005version} are
essential tools for managing changes to source code over time.
They facilitate collaboration, track modifications,
and enable reversion to earlier versions when necessary.
VCSs play a vital role in modern software development by
ensuring that all changes are systematically recorded and maintained.
As a result, the documentation of code changes naturally
includes the introduction of bugs.
Bisect is a powerful feature supported by many VCSs---most notably
in Git~\cite{gitbisect}---that allows developers to efficiently
identify the specific commit in which a bug or issue was introduced.
By automating a binary search through the commit history,
bisection significantly streamlines the debugging process.
While bisection is widely adopted and studied for debugging purposes,
its potential application in bug deduplication remains
largely overlooked and underexplored in the existing literature.

\subsection{Motivation}
The limitations explained in \cref{subsec:existing-techniques}
inevitably restrict the practicality of \emph{analysis-based}
deduplication approaches in real-world scenarios.
This highlights the practical need for a more generalizable,
easy-to-implement, and less-dependent approach.
Bisection, given its high simplicity and effectiveness,
motivates us to conduct this study to explore
its potential for compiler bug deduplication.
The fundamental insight driving this approach is that test failures
introduced by different commits should correspond to different root
causes, \ie, distinct bugs.

\emph{
    By exploring development history through bisection,
    our study aims to provide a simple, generalizable,
    and effective perspective on bug deduplication,
    reducing the complexity and overhead associated
    with existing analysis-based techniques.
}

\section{Effectiveness}
\label{sec:effectiveness}

Effectiveness is of paramount importance in compiler
bug deduplication, as it directly impacts the practical
utility of the deduplication approach by determining
how much human effort can be saved in debugging.
To explore this, we pose two effectiveness-oriented research questions:
\begin{itemize} [leftmargin=*]
    \item \textbf{RQ1 (Bisection as the Sole Criterion):}
    \rqOne

    \item \textbf{RQ2 (False Negative Mitigation):}
    \rqTwo
\end{itemize}

By answering these questions, we aim to demonstrate the feasibility of
deduplicating compiler bugs with bisection and provide practical guidelines
for its application in real-world scenarios.

\subsection{Datasets}

In this study, we use five datasets.
Four are drawn from prior work on compiler bug deduplication~\cite{tamer,d3}
and compiler testing~\cite{chen2022boosting,chen2019history},
while the fifth is newly constructed for this study.
When constructing these benchmark suites,
we strive to reproduce each test failure within our
experimental environment, while excluding test programs
that meet either of the following criteria:
(1) they exhibit undefined behaviors (UBs), as compilers
are not designed to produce consistent outputs for such cases
and may make arbitrary decisions---such as applying
optimizations---around UBs; or
(2) the associated bugs could not be reliably reproduced
in our experimental environment.
The latter exclusion primarily results from differences in system
configurations or architectures
and affects only a minimal number of cases.

\begin{table}[h]
    \caption{
        The number of test programs and unique bugs in each dataset.
    }
    \label{tab:benchmark-info}
    \setlength{\tabcolsep}{17pt}
    \resizebox{0.6\columnwidth}{!}{%
    \begin{tabular}{@{}ccc@{}}
    \toprule
    \quad Dataset  & Number of Programs & Number of Bugs\;\quad \\ \midrule
    \quad GCC-4.3.0  & \benchmarkGCCFourThreeZeroSize        & \benchmarkGCCFourThreeZeroBugs\quad             \\
    \quad GCC-4.4.0  & \benchmarkGCCFourFourZeroSize         & \benchmarkGCCFourFourZeroBugs\quad              \\
    \quad GCC-4.5.0  & \benchmarkGCCFourFiveZeroSize         & \benchmarkGCCFourFiveZeroBugs\quad              \\
    \quad LLVM-2.8.0 & \benchmarkLLVMTwoEightZeroSize        & \benchmarkLLVMTwoEightZeroBugs\quad             \\
    \quad GCC-13.1.0 & \benchmarkGCCThirdteenSize            & \benchmarkGCCThirdteenBugs\quad                 \\ \bottomrule
    \end{tabular}%
    }
    \end{table}

Through considerable effort, we constructed five well-organized datasets.
As summarized in \cref{tab:benchmark-info},
each dataset comprises a collection of bug-triggering test programs
associated with a specific compiler version.
The first four datasets,
namely \gcc-4.3.0, \gcc-4.4.0, \gcc-4.5.0, and \llvm-2.8.0,
cover all publicly available benchmarks that have been
used in prior work on compiler bug deduplication,
thereby ensuring both comparability and comprehensiveness.

Although comprehensive, these datasets are relatively old,
involving compiler versions released more than a decade ago.
To further validate the generalizability of our approach in
modern compiler development settings, we constructed a new
dataset for a recent compiler version, \ie, \gcc-13.1.0.
We built this dataset as follows.
To obtain a set of distinct bug-triggering programs,
we first used \legofuzz~\cite{ni2025legofuzz},
a recently proposed compiler fuzzing framework,
to stress test \gcc-13.1.0.
Consistent with the scope of this study,
we focused exclusively on silent miscompilation bugs.
During a two-week fuzzing campaign, \legofuzz generated
\benchmarkGCCThirdteenSize programs that triggered
miscompilation bugs in \gcc-13.1.0.
To determine the ground truth, \ie,
the unique bugs triggered by these programs,
we adopted the correcting-commit method used
in prior work by \tamer~\cite{tamer}.
Specifically, for each bug-triggering program,
we first identified a good version later than
\gcc-13.1.0, such as \gcc-14.1.0,
that executes the program correctly.
We then performed bisection between \gcc-13.1.0
and the good version to identify the first commit
that fixed the bug and caused the program to execute correctly.
Programs that shared the same correcting commit were
then grouped as triggering the same bug.
Note that this method cannot be used to
deduplicate new bugs without known patches,
thus does not threaten the validity of our study.
Using this method, we constructed a dataset of
\benchmarkGCCThirdteenSize test programs that trigger
\benchmarkGCCThirdteenBugs unique miscompilation bugs in \gcc-13.1.0.

Importantly, all bugs and bug-triggering programs
in these datasets were originally collected
from real-world compiler testing
efforts~\cite{chen2022boosting,chen2019history,csmith,ni2025legofuzz},
which ensures the practical relevance of our study.
The datasets feature two major compilers, GCC and LLVM,
both of which have long and complex development histories,
allowing the datasets to faithfully reflect the intricacies
of real-world compiler development and testing scenarios.
The high ratio of test programs to unique bugs in each dataset
underlines the practical challenge of
compiler bug deduplication in realistic settings.
Among the five datasets,
the \gcc-4.3.0 dataset stands out as the largest and most complex,
representing the most challenging case for deduplication.

\subsection{Bisection as the Sole Criterion}
In the first place, we investigate how effective the results of bisection,
\ie, the failure-inducing commits, serve as the sole criterion for
compiler bug deduplication. This examination establishes
the foundation for our subsequent study.

\subsubsection{Methodology}
\label{subsubsection:rq1-methodology}
We follow the workflow of existing bug deduplication research~\cite{tamer,d3,transformer},
which prioritizes test programs based on customized distance metrics.
Specifically, our bisection-based deduplication approach
involves the following steps:

\noindent\underline{\emph{Step 1 (Bisection Range Initialization):}}
For each test program \testInput in the dataset,
we initialize a bisection range with a known-good version
(\goodVersion) and a known-buggy version (\badVersion).
\badVersion is obviously the compiler version exhibiting the bug,
\eg, GCC-4.3.0.
To quickly determine \goodVersion,
we build a series of main releases of the target compiler
(\eg, GCC-4.2.0, GCC-4.1.0, etc.).
Then we check the behavior of \testInput with each of these
versions, and set \goodVersion to the first version before \badVersion
where the bug no longer manifests with \testInput.
We maintain a list of pre-built compiler binaries to facilitate
quickly narrowing down the bisection range.

\noindent\underline{\emph{Step 2 (Failure-inducing Commit Localization):}}
For each program \testInput in the dataset,
we perform bisection, technically a binary search,
within the identified range (\goodVersion to \badVersion)
on the compiler's commit history
to locate the earliest commit \commit that induces the failure for
\testInput.

\noindent\underline{\emph{Step 3 (Distance Calculation):}}
The distance between two test programs reflects their estimated
similarity in terms of bug triggering. In our approach,
we calculate the distance between two programs based on the time
interval between their corresponding failure-inducing commits.
Given two test programs \testInputAlpha and \testInputBeta,
with their failure-inducing commits \commitAlpha and \commitBeta,
the distance $\mathcal{D}$ between them is defined as:
\begin{equation} 
    \mathcal{D}(\testInputAlpha, \testInputBeta) =
    \mathcal{D}_{\textit{bisect}}(\commitAlpha, \commitBeta) =
    |t(\commitAlpha) - t(\commitBeta)|
\label{eq:distance-bisect}
\end{equation}
where $t(\commit)$ denotes the timestamp of the commit $\commit$
in integer seconds.
The underlying rationale
is that a larger time interval indicates greater likelihood that the
test programs correspond to distinct bugs.

\noindent\underline{\emph{Step 4 (Test Program Prioritization):}}
Prioritization aims to order test programs such that those
triggering distinct bugs appear earlier,
thus facilitating faster identification of unique bugs.
Using the calculated distances, we prioritize test programs according
to the \emph{furthest point first} (FPF) algorithm~\cite{gonzalez1985clustering},
which always selects the next element
that is farthest from all previously chosen ones,
thereby ensuring that more diverse programs appear earlier in the ranked list.
Specifically, we start with an empty list \rankedList, and
randomly select the initial test program to append to \rankedList.
Then, for each remaining program, we compute its distances to all
programs already in \rankedList, and record the \emph{minimum distance}.
The program with the \emph{largest minimum distance} is selected
next and appended to \rankedList.
This process continues until \rankedList contains all the test programs.
The resulting ranked list then serves as the guidance for
the bug reporting and fixing process,
allowing developers to efficiently identify and
resolve diverse bugs while reviewing fewer programs,
significantly reducing their debugging effort.

\subsubsection{Implementation}
We automated the bisection-based bug deduplication approach using
\texttt{git bisect}~\cite{gitbisect},
a built-in Git functionality,
together with a set of helper scripts.
Specifically, \texttt{git bisect} performs binary search automatically
within the compiler repository and,
at each step, invokes a predefined test script
to determine whether the current compiler version
is good or bad for the given test program.
This script builds the compiler at the current commit,
compiles and executes the test program
with the resulting compiler,
and returns 0 or 1 to indicate whether the behavior is good or bad.
To further accelerate the process,
we took several additional measures.
First, we maintained a list of prebuilt compiler
binaries to quickly narrow the bisection range.
Specifically, we built a series of major releases
of the target compiler.
For example, when deduplicating bugs in \gcc-4.3.0,
we built several earlier releases,
such as \gcc-4.2.0 and \gcc-4.1.0.
For each test case,
we then checked the program's behavior on
these versions from newest to oldest and selected
as the good version the latest one before the bad version
in which the bug no longer manifests.
This substantially narrowed the bisection range
and reduced the cost of bisection.
In addition,
compiler versions built during bisection were
cached and reused in subsequent bisection jobs,
further reducing time and resource consumption.
As shown in \cref{sec:efficiency},
this caching mechanism significantly reduces
the total number of required builds and
thereby improves efficiency.

Because some compilers in our datasets are relatively old and may be incompatible
with modern system configurations, we adopted a container-based setup to provide
a consistent and controlled environment for the bisection process.
In particular, we ran the process inside an Ubuntu 12.04 container,
whose software environment is better suited to the targeted compiler versions.
This container-based setup was required only for our study because of the age of the datasets;
it does not limit the applicability of the approach in real-world settings,
where the compilers under test are typically much more recent.

\subsubsection{Compared Techniques}
In this study, we compare the bisection-based deduplication
approach with two state-of-the-art analysis-based techniques
introduced in \cref{subsec:existing-techniques},
namely \tamer~\cite{tamer} and \dthree~\cite{d3}.
For \tamer, we adopt the compiler function coverage achieved
by the bug-triggering test program, \ie, the number of execution
times for each function, as the distance metric,
which is shown to be the most effective metric~\cite{tamer}
and has been adopted by prior work for evaluation~\cite{d3}.
For \dthree, we follow the default configurations provided in the original paper.
We do not include \transformer~\cite{transformer} in our comparison,
as its scope is limited to transformation-based
compiler testing. Moreover, prior work~\cite{d3} has demonstrated
that its performance degrades significantly when applied to
general compiler testing scenarios.

\subsubsection{Metrics}
\label{subsec:rq1-metrics}
The primary goal of bug deduplication is enabling developers to
uncover the maximum number of distinct bugs by examining as few
test programs as possible.
As this research question serves as the initial exploration,
we adopt the intuitive metric of a \emph{bug discovery curve}
for preliminary evaluation of effectiveness.
A bug discovery curve visually illustrates how efficiently a ranked list
enables the examination of the test cases one by one to encounter
at least one representative from each distinct bug
category~\cite{pelleg2004active,vatturi2009category}.
Thus, a steeper curve is preferable, indicating greater deduplication
effectiveness.
The bug discovery curve provides a visually intuitive assessment of whether
bisection alone effectively functions as a bug deduplication criterion.
For a more quantitative evaluation at a large scale, we later adopt
the metric of \emph{wasted effort},
which calculates the number of test cases examined before identifying
each distinct bug.

\begin{figure*}[h!]
    \centering
    \begin{subfigure}[b]{0.49\columnwidth}
        \includegraphics[width=\columnwidth]{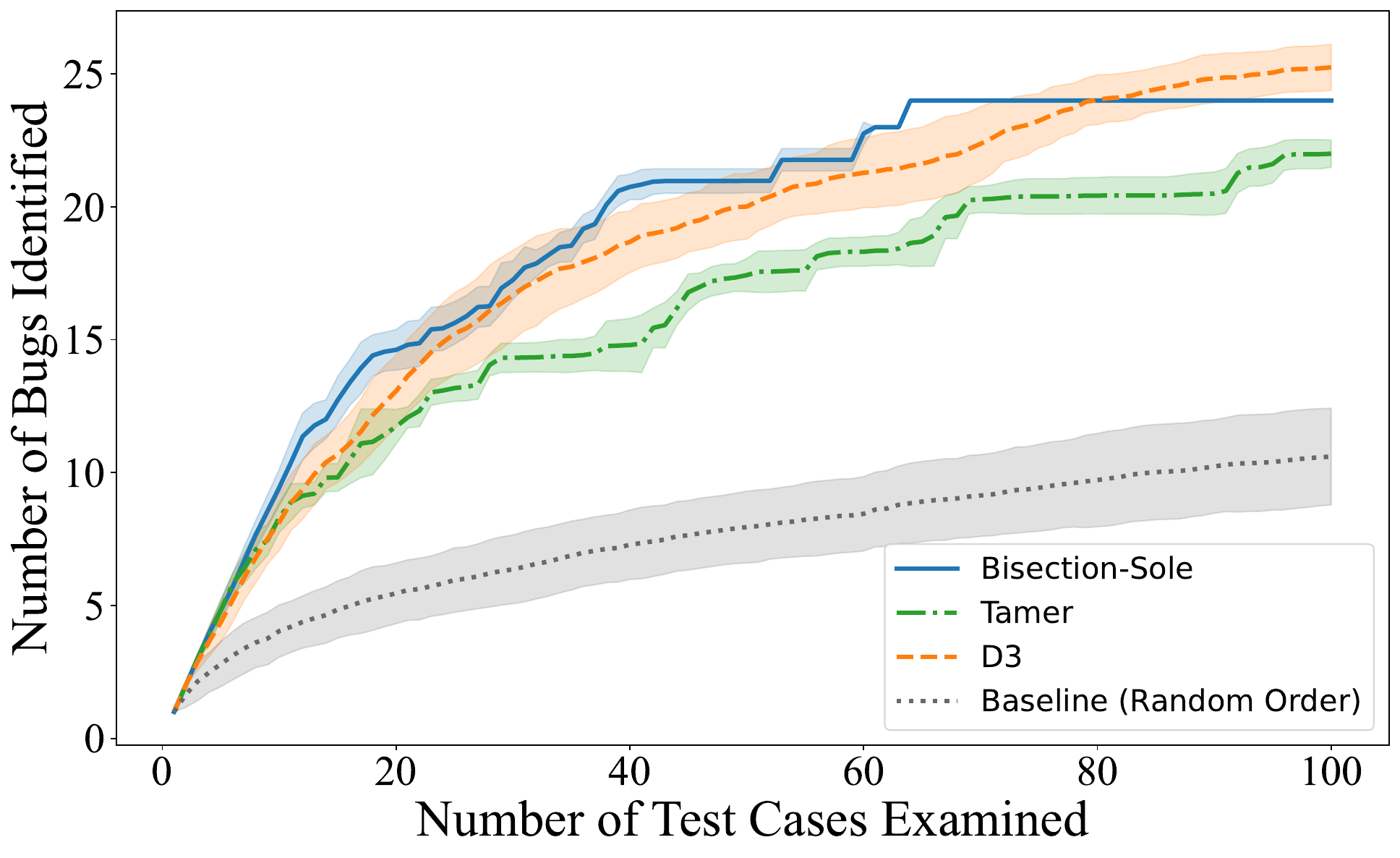}
       \caption{
       	    First 100 tests
       	}
       \label{subfig:rq1-gcc430-curve-first-100}
    \end{subfigure}
        \hfil
    \begin{subfigure}[b]{0.49\columnwidth}
        \includegraphics[width=\columnwidth]{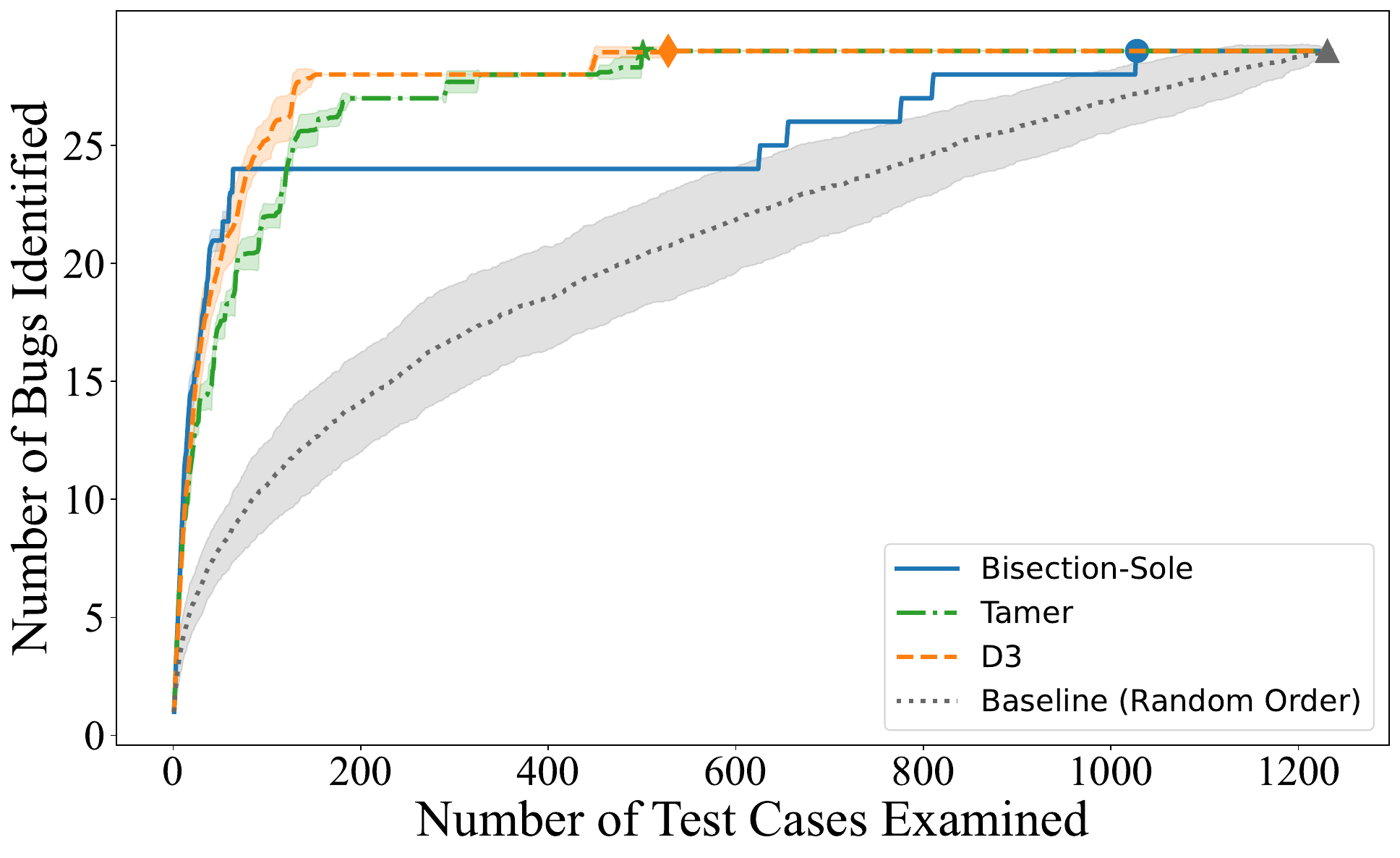}
        \caption{
            All tests
        }
        \label{subfig:rq1-gcc430-curve-all-tests}
    \end{subfigure}
    \caption{
        Bug discovery curves of \bisectionSole, \tamer, \dthree and the baseline on the \gcc-4.3.0 dataset.
    }
    \label{fig:rq1-gcc430-curve}
\end{figure*}

\subsubsection{Results and Analysis}
\label{subsec:rq1-results}
To answer RQ1, we perform the bisection-based deduplication
approach described in \cref{subsubsection:rq1-methodology}
on the \gcc-4.3.0 dataset, selected
for its largest size and highest complexity among the five datasets.
To alleviate the impact of randomness in the prioritization process,
each deduplication approach was repeated one hundred times,
and we report the averaged results.
The bug discovery curves of the three evaluated techniques, namely
\tamer, \dthree, and \bisectionSole, are presented in \cref{fig:rq1-gcc430-curve},
along with a random ordering baseline.
The shaded region represents one standard deviation above and below the mean.
\cref{subfig:rq1-gcc430-curve-first-100}
specifically shows the results for the first 100 test cases examined,
reflecting common developer practices where not all bug reports would
be thoroughly examined~\cite{chen2023exploring,tamer,d3,holmes2018causal}.
\cref{subfig:rq1-gcc430-curve-all-tests} provides a view of the entire
deduplication process.
Overall, the results shown in these curves are encouraging,
demonstrating the technical feasibility of using bisection
as a criterion for deduplicating compiler bugs.
Specifically, as depicted in \cref{subfig:rq1-gcc430-curve-first-100},
\bisectionSole successfully identifies 24 unique bugs within the first
100 test programs examined, which is more than the 22 bugs identified
by \tamer and slightly fewer than the 25 bugs identified by \dthree.
Given the inherent simplicity of bisection, 
these findings offer promising support for the practical potential
of using bisection for compiler bug deduplication.

However, the limitation of using such a single-criterion
strategy is also apparent.
As shown in \cref{subfig:rq1-gcc430-curve-all-tests},
after identifying 24 bugs by examining 64 test programs,
the discovery curve of \bisectionSole exhibits several
prolonged plateaus, significantly slowing the identification
of new bugs. Ultimately, \bisectionSole requires examining an
average of \bisectionSoleFindAllBugs test programs to uncover all
\benchmarkGCCFourThreeZeroBugs bugs in the dataset, which is
substantially more than the \tamerFindAllBugs and \dthreeFindAllBugs
cases required by \tamer and \dthree, respectively, and only slightly
fewer than the random baseline.
This low effectiveness is primarily attributed to the
misclassifications in the deduplication results.
These misclassifications can be categorized into two types:

(1) \emph{\textbf{False positives}}, where programs triggering the
same underlying bug are incorrectly classified as
distinct due to different bisection results.

(2) \emph{\textbf{False negatives}}, where programs triggering
distinct bugs are incorrectly grouped together
due to the same bisection result.

False positives naturally arise because a single bug may manifest
in various ways, leading to different failure-inducing commits
(a failure-inducing commit is not necessarily the root cause
of the bug, which may simply make the bug triggerable by a program).
Given this inherent nature of compiler bugs,
false positives are not unique to our bisection-based approach,
but prevail in other deduplication methods as well.
For instance, for analysis-based approaches,
false positives occur when two test programs
triggering the same bug are classified as distinct
due to different analysis results, \ie,
different execution paths or different covered code regions.
The direct consequence of false positives is the generation of
duplicate bug reports, which require manual effort from
developers to filter out.

Conversely, false negatives pose a more serious problem,
as they directly result in overlooked bugs unless all test cases
are examined---contrary to the fundamental goal of bug deduplication,
and often impractical in real-world scenarios~\cite{tamer, chen2023exploring, holmes2018causal}.
In our approach which uses bisection results as the sole criterion,
false negatives typically stem from scenarios such as large single
commits introducing multiple different bugs, which are frequent
occurrences in compiler development. In other cases, commits modifying
configuration flags (\eg, macro definitions) might disclose multiple bugs
across different compiler components. Instead of manually investigating
each occurrence, developing an effective yet lightweight strategy to
mitigate such false negatives is crucial, forming the core focus of RQ2.

\myFinding{
    Employing bisection as the criteria to deduplicate compiler bugs
    is technically feasible, as demonstrated by its effectiveness in
    identifying the majority of bugs within the \gcc-4.3.0 dataset.
    However, the overall performance and practical utility could be
    further improved with extra strategies to mitigate the
    occurrence of false negatives.
}

\subsection{RQ2: False Negative Mitigation}
\label{subsec:rq2}

The presence of false negatives poses a practical limitation to the
utility of bisection-based bug deduplication. In this research question,
we investigate strategies to mitigate this issue and improve the
overall effectiveness of bug deduplication using bisection.
To preserve the inherent simplicity and efficiency of bisection,
any additional measures introduced to mitigate false negatives must
remain lightweight, avoiding complex, analysis-based methods.
Within this context,
examining various compiler modes and configurations that trigger bugs
across different programs provides a promising direction.
In particular, given that modern compilers typically support
customizable optimization settings---and that complex optimizations
are frequently a source of compiler bugs~\cite{emi, csmith}---exploring
these optimizations as potential bug-triggering factors
aligns well with our goal of developing a simple yet
effective solution.

In compiler testing, test programs are typically
executed under various optimization levels, and bugs may
manifest exclusively under specific optimizations are enabled.
Common optimization flags in GCC and LLVM, such as \texttt{-O1},
\texttt{-O2}, \texttt{-O3}, and \texttt{-Os}, enable predefined
sets of optimizations that can influence bug manifestation.
However, not all optimizations enabled by these flags are directly
responsible for triggering the bugs. For example, two test programs
might both trigger bugs under \texttt{-O3}, but one bug could be
caused specifically by the optimization \texttt{-tree-fre},
while the other might result from \texttt{-inline-small-functions}.
Thus, pinpointing the exact optimizations responsible for triggering each bug is crucial.

Previous work~\cite{d3} leverages the idea of Delta Debugging~\cite{deltadebugging}
to identify the minimal optimization set that still trigger a given bug.
While this approach is effective in principle,
it faces a practical limitation:
some basic optimizations within predefined optimization levels
(e.g., \texttt{-O1}, \texttt{-O2}) are implicitly enabled
and cannot be directly controlled via command line flags.
These implicit optimizations are sometimes the precondition
for triggering a bug.
For example, in our datasets,
some test programs trigger bugs under \texttt{-O1},
yet the bugs persist even after all explicitly optimizations
associated with \texttt{-O1} are disabled.
Conversely, enabling these optimizations individually
without using \texttt{-O1} fails to reproduce the bug.
Identifying such implicit optimizations typically
requires in-depth examination of the compiler source
code---an effort-intensive process that conflicts with
our goal of maintaining simplicity.
To resolve this issue, we propose \emph{reversing} the original strategy.
Rather than explicitly enabling optimization flags to trigger a bug,
we retain the original optimization level (e.g., \texttt{-O1}),
and selectively add flags to \emph{disable} optimizations outside
the minimal triggering set.
This reversed approach allows us to isolate critical optimizations
without the need for source-level analysis. The resulting optimization
configuration is then integrated with the bisection outcomes
to form a combined deduplication criterion.

\subsubsection{Methodology}

The overall deduplication process remains consistent with that
described in RQ1, except we introduce an additional step to extract
optimization information for false negative mitigation,
and modify the distance calculation step accordingly.

\noindent\underline{\emph{Extra Step (Optimization Information Extraction):}}
For each test program \testInput in the dataset,
suppose it initially triggers a bug under a certain optimization level,
denoted generically as \optLevelOfInputI.
In our datasets, \optLevelOfInputI is one of the standard optimization
levels defined by the \gcc and \llvm compilers,
\ie, \texttt{O0}, \texttt{O1}, \texttt{O2}, \texttt{O3}, and \texttt{Os}.
Each optimization level explicitly enables a set of optimization passes
$\optLevelOfInputI \coloneq \{ o_1, o_2,..., o_n \}$.
We then apply the minimizing Delta Debugging (\ddmin) algorithm~\cite{deltadebugging}
to identify the minimal set of optimization passes
that is necessary to trigger the bug with \testInput.
Generally, \ddmin is a divide-and-conquer approach that iteratively
partitions a set into smaller subsets, identifying and removing
unnecessary items.
Specifically, \ddmin first splits the list of passes into
$n = 2$ partitions. Them, it traverses each partition
and checks whether these is a partition that can solely
trigger the bug with \testInput. If so, it removes all
other partitions and starts over with the remaining
partition. Otherwise, it traverses each partition again
to check whether the complement of any partition can
trigger the bug with \testInput. If so, it removes
the partition and starts over with the remaining
partitions. If the list cannot be reduced during
the above process, \ddmin further splits the list
into $2n$ partitions and repeats the above steps.
Throughout the process of \ddmin,
when checking the bug-triggering with a subset of optimizations passes,
we always keep the original optimization level
$\optLevelOfInputI$ in execution,
and selectively disable optimization passes outside the subset
by adding disabling flags
(formatted as \texttt{-fno-\{pass\_name\}} in \gcc and \llvm)
to the compiler command line.
This approach ensures that basic optimizations inherently tied to
the optimization level remain active, eliminating the need for manual
inspection of the compiler's internal implementation.
Eventually, we obtain a minimal set of optimization passes,
$\optLevelOfInputI' = \{ o_1, o_2,..., o_m \} \subseteq \optLevelOfInputI$,
where $m \leq n$,
that are necessary to reproduce the bug for the given test program \testInput.

\begin{figure*}[h!]
    \centering
    \begin{subfigure}[b]{0.48\columnwidth}
        \includegraphics[width=\columnwidth]{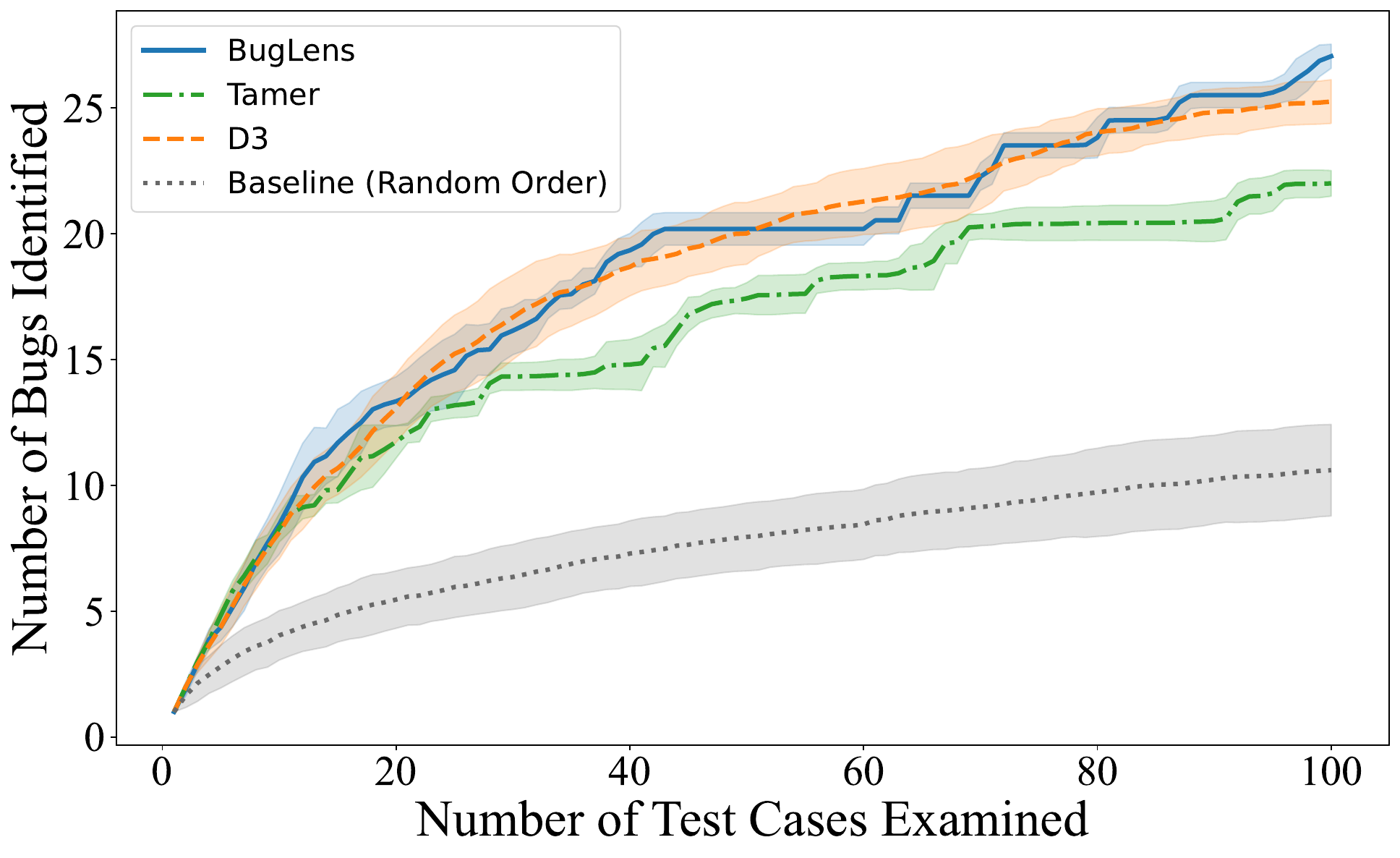}
        \caption{
       	    First 100 tests
       	}
       \label{subfig:rq2-gcc430-curve-first-100}
    \end{subfigure}
        \hfil
    \begin{subfigure}[b]{0.48\columnwidth}
        \includegraphics[width=\columnwidth]{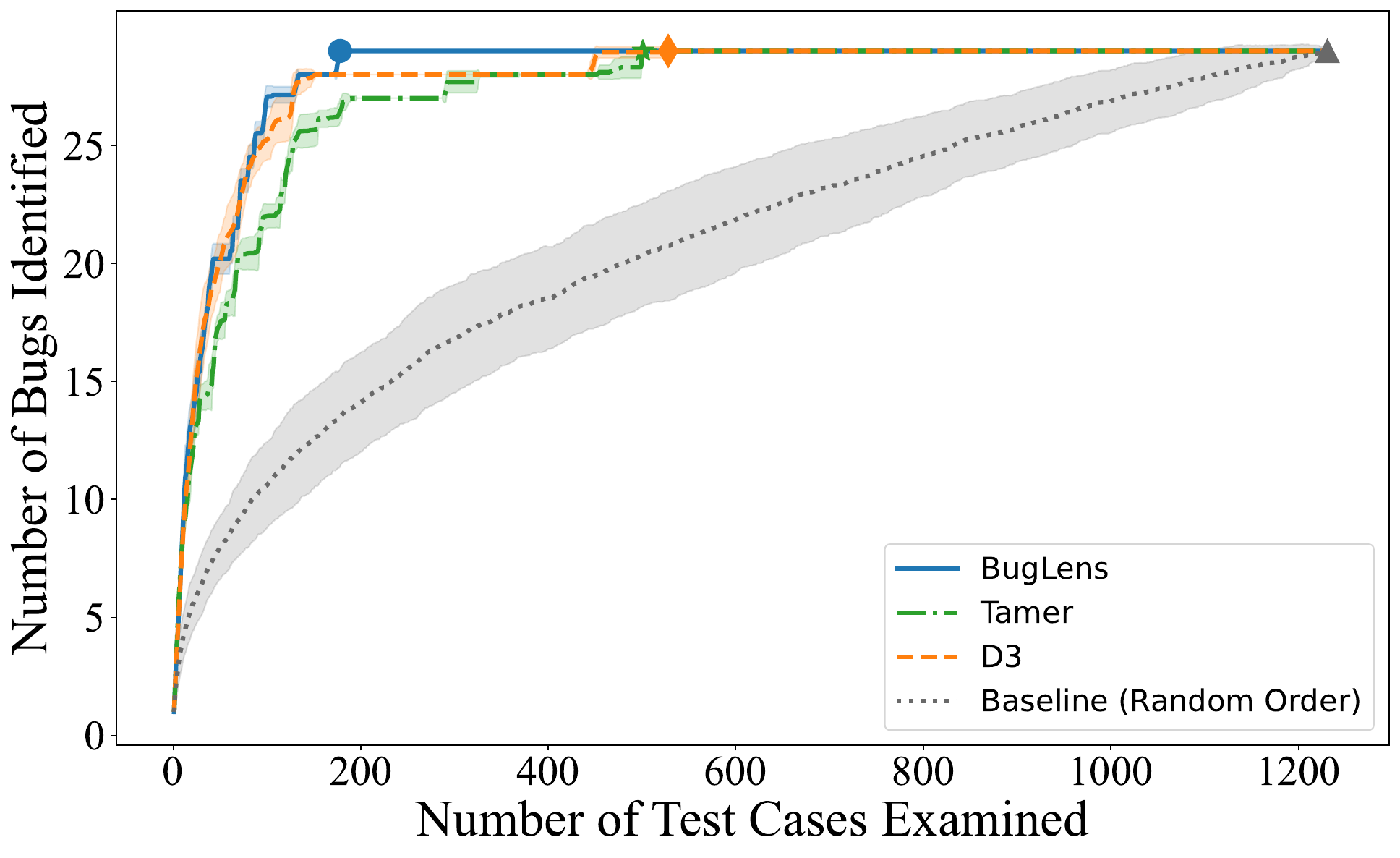}
        \caption{
            All tests
        }
        \label{subfig:rq2-gcc430-curve-all-tests}
    \end{subfigure}
    \caption{
        Bug discovery curves of \proj, \tamer, \dthree and the baseline on the \gcc-4.3.0 dataset.
    }
    \label{fig:rq2-gcc430-curve}
\end{figure*}

\noindent\underline{\emph{Revised Step 3 (Distance Calculation):}}
To calculate distances between test programs, we construct
vectors that represent their respective bug-triggering configurations.
Specifically, each test program is encoded as a vector
$\mathbf{v} = [v_0, v_1,..., v_{n}]$,
where $n$ is the total number of optimization
passes supported by the compiler.
The first element $v_0 \in \{\texttt{O0}, \texttt{O1}, \texttt{O2}, \texttt{O3}, \texttt{Os}\}$
denotes the optimization level under which the bug manifests.
The remaining elements $v_1, v_2,..., v_{n}$ are binary values
indicating whether the corresponding optimization pass is included
in the minimal bug-triggering set $\optLevelOfInputI'$.
Specifically, for $i \in [1, n]$,
$v_i = 1$ if the corresponding optimization pass is part of
$\optLevelOfInputI'$, and $0$ otherwise.
Then, the distance between two vectors
$\mathbf{v}^{\alpha} = [v_{0}^{\alpha}, v_{1}^{\alpha},..., v_{n}^{\alpha}]$ and
$\mathbf{v}^{\beta} = [v_{0}^{\beta}, v_{1}^{\beta},..., v_{n}^{\beta}]$ is computed as follows:
\begin{equation}
    \mathcal{D}_\textit{opt}(\mathbf{v}^{\alpha}, \mathbf{v}^{\beta}) =
    \frac{1}{n+1}\sum_{k=0}^{n}\mathbb{I}(v^{\alpha}_k \neq v^{\beta}_k)
\end{equation}
where $\mathbb{I}(x)$ is the indicator function, returning 1 if its
argument is true and 0 otherwise.
The distance value is normalized to the range $[0, 1]$ and serves as a
complementary criterion to the original distance measure based on the
time interval between failure-inducing commits.
Thus, the revised distance $\mathcal{D}$ between two test programs
\testInputAlpha (with bug-inducing commit $\commitAlpha$ and optimization vector $\vectorAlpha$)
and \testInputBeta (with bug-inducing commit $\commitBeta$ and optimization vector $\vectorBeta$)
is defined as:
\begin{equation}
    \mathcal{D}(\testInputAlpha, \testInputBeta) =
    \mathcal{D}_{\textit{bisect}}(\commitAlpha, \commitBeta) +
    \mathcal{D}_{\textit{opt}}(\vectorAlpha, \vectorBeta)
\end{equation}
where $\mathcal{D}_{\textit{bisect}}(\commitAlpha, \commitBeta)$
is the original distance measure defined in \cref{eq:distance-bisect},
computed as the time interval between the failure-inducing
commits $\commitAlpha$ and $\commitBeta$.
This revised distance metric allows us to distinguish between
bugs that cannot be differentiated by bisection alone,
thereby reducing false negatives.

The remaining steps of the deduplication process
follow the same procedure as in RQ1.
Test programs are prioritized using the combined distance measure
in a furthest-point-first (FPF) order, ensuring that
those triggering distinct bugs appear earlier in the list.
This ranked list then serves as a strong guide for bug reporting
and resolution, potentially saving substantial developer effort.
We refer to this refined approach as \proj.

\subsubsection{Results and Analysis}

We maintain the same experimental settings from RQ1,
and perform bug deduplication with \proj, \tamer, and \dthree
on the \gcc-4.3.0 dataset.
The bug discovery curves are shown in \cref{fig:rq2-gcc430-curve}.
As illustrated in \cref{subfig:rq2-gcc430-curve-first-100},
during the early stage,
\proj continues to perform competitively with
\dthree and still outperforms \tamer.
Specifically, during the first 100 test case examinations,
\proj successfully identifies 27 distinct bugs,
while \dthree and \tamer identify 25 and 22 bugs, respectively,
indicating that \proj is slightly more effective than the
analysis-based approaches in the early stages.
The benefit of incorporating
optimization information as a complementary criterion
becomes evident in the middle-to-late stages,
where it successfully distinguishes cases that could not be
differentiated solely by bisection results.
Specifically, as presented in \cref{subfig:rq2-gcc430-curve-all-tests},
with \proj,
only \bisectionWithOptFindAllBugs test case examinations are required
on average to identify all \benchmarkGCCFourThreeZeroBugs bugs in the dataset,
which is a substantial reduction of \bisectionOptFindAllBugsSaveEffortThanTamer
compared to \tamer (\tamerFindAllBugs)
and \bisectionOptFindAllBugsSaveEffortThanDthree
compared to \dthree (\dthreeFindAllBugs).
Notably, compared to \bisectionSole, which requires
examining \bisectionSoleFindAllBugs test cases on average
to identify all bugs, \proj reduces the number of
by 82.9\%. These results demonstrate
leveraging specific bug-triggering optimizations provides an effective
strategy for mitigating false negatives in compiler bug deduplication with bisection.

\myFinding{
    Exploring bug-triggering optimizations offers an
    promising way to reducing false negatives in bug deduplication
    while requiring only a modest extra effort.
    \proj then significantly outperforms
    the other evaluated techniques in overall effectiveness,
    by saving \bisectionOptFindAllBugsSaveEffortThanTamer and
    \bisectionOptFindAllBugsSaveEffortThanDthree of the wasted effort
    compared to \tamer and \dthree, respectively.
    Given the simplicity of techniques involved,
    \proj offers a highly practical solution
    for compiler bug deduplication in real-world development and testing.
}

\subsection{Large Scale Evaluation}
\label{subsec:large-scale-evaluation}

\begin{table*}[]
    \centering
    \caption{
        Comparison of wasted effort between \proj and other analysis-based techniques.
    }
    \label{tab:waste-effort}
    \setlength{\tabcolsep}{7pt}
    \resizebox{\textwidth}{!}{%
    \begin{tabular}{crrrrrcrrrrr}
        \toprule
        \multicolumn{12}{c}{\textbf{GCC-4.3.0}} \\ \midrule
        ID & \cellcolor[HTML]{EBEBEB}\proj & \tamer & \cellcolor[HTML]{EBEBEB}$\Delta_{\%}(\tamer)$ & \dthree & \multicolumn{1}{r|}{\cellcolor[HTML]{EBEBEB}$\Delta_{\%}(\dthree)$} & ID & \cellcolor[HTML]{EBEBEB}\proj & \tamer & \cellcolor[HTML]{EBEBEB}$\Delta_{\%}(\tamer)$ & \dthree & \cellcolor[HTML]{EBEBEB}$\Delta_{\%}(\dthree)$ \\ \midrule
        2  & \cellcolor[HTML]{EBEBEB}2.06   & 2.0    & \cellcolor[HTML]{EBEBEB}-3.00\%  & 2.0    & \multicolumn{1}{r|}{\cellcolor[HTML]{EBEBEB}-3.00\%}  & 16 & \cellcolor[HTML]{EBEBEB}28.05  & 41.67  & \cellcolor[HTML]{EBEBEB}32.69\% & 27.24  & \cellcolor[HTML]{EFEFEF}-2.97\%  \\
        3  & \cellcolor[HTML]{EBEBEB}3.08   & 3.0    & \cellcolor[HTML]{EBEBEB}-2.67\%  & 3.16   & \multicolumn{1}{r|}{\cellcolor[HTML]{EBEBEB}2.53\%}   & 17 & \cellcolor[HTML]{EBEBEB}30.76  & 44.93  & \cellcolor[HTML]{EBEBEB}31.54\% & 30.11  & \cellcolor[HTML]{EFEFEF}-2.16\%  \\
        4  & \cellcolor[HTML]{EBEBEB}4.08   & 4.16   & \cellcolor[HTML]{EBEBEB}1.92\%   & 4.46   & \multicolumn{1}{r|}{\cellcolor[HTML]{EBEBEB}8.52\%}   & 18 & \cellcolor[HTML]{EBEBEB}34.69  & 52.74  & \cellcolor[HTML]{EBEBEB}34.22\% & 34.01  & \cellcolor[HTML]{EFEFEF}-2.00\%  \\
        5  & \cellcolor[HTML]{EBEBEB}5.71   & 5.16   & \cellcolor[HTML]{EBEBEB}-10.66\% & 5.75   & \multicolumn{1}{r|}{\cellcolor[HTML]{EBEBEB}0.70\%}   & 19 & \cellcolor[HTML]{EBEBEB}38.12  & 62.13  & \cellcolor[HTML]{EBEBEB}38.64\% & 38.99  & \cellcolor[HTML]{EFEFEF}2.23\%   \\
        6  & \cellcolor[HTML]{EBEBEB}6.92   & 6.24   & \cellcolor[HTML]{EBEBEB}-10.90\% & 6.94   & \multicolumn{1}{r|}{\cellcolor[HTML]{EBEBEB}0.29\%}   & 20 & \cellcolor[HTML]{EBEBEB}42.57  & 67.47  & \cellcolor[HTML]{EBEBEB}36.91\% & 48.35  & \cellcolor[HTML]{EFEFEF}11.95\%  \\
        7  & \cellcolor[HTML]{EBEBEB}7.96   & 7.77   & \cellcolor[HTML]{EBEBEB}-2.45\%  & 8.32   & \multicolumn{1}{r|}{\cellcolor[HTML]{EBEBEB}4.33\%}   & 21 & \cellcolor[HTML]{EBEBEB}57.78  & 83.95  & \cellcolor[HTML]{EBEBEB}31.17\% & 54.86  & \cellcolor[HTML]{EFEFEF}-5.32\%  \\
        8  & \cellcolor[HTML]{EBEBEB}9.35   & 9.34   & \cellcolor[HTML]{EBEBEB}-0.11\%  & 9.59   & \multicolumn{1}{r|}{\cellcolor[HTML]{EBEBEB}2.50\%}   & 22 & \cellcolor[HTML]{EBEBEB}67.09  & 94.89  & \cellcolor[HTML]{EBEBEB}29.30\% & 61.76  & \cellcolor[HTML]{EFEFEF}-8.63\%  \\
        9  & \cellcolor[HTML]{EBEBEB}10.57  & 10.88  & \cellcolor[HTML]{EBEBEB}2.85\%   & 11.12  & \multicolumn{1}{r|}{\cellcolor[HTML]{EBEBEB}4.95\%}   & 23 & \cellcolor[HTML]{EBEBEB}71.18  & 112.25 & \cellcolor[HTML]{EBEBEB}36.59\% & 68.4   & \cellcolor[HTML]{EFEFEF}-4.06\%  \\
        10 & \cellcolor[HTML]{EBEBEB}11.76  & 14.52  & \cellcolor[HTML]{EBEBEB}19.01\%  & 12.94  & \multicolumn{1}{r|}{\cellcolor[HTML]{EBEBEB}9.12\%}   & 24 & \cellcolor[HTML]{EBEBEB}77.01  & 119.29 & \cellcolor[HTML]{EBEBEB}35.44\% & 76.5   & \cellcolor[HTML]{EFEFEF}-0.67\%  \\
        11 & \cellcolor[HTML]{EBEBEB}13.24  & 17.22  & \cellcolor[HTML]{EBEBEB}23.11\%  & 15.08  & \multicolumn{1}{r|}{\cellcolor[HTML]{EBEBEB}12.20\%}  & 25 & \cellcolor[HTML]{EBEBEB}84.15  & 125.72 & \cellcolor[HTML]{EBEBEB}33.07\% & 91.08  & \cellcolor[HTML]{EFEFEF}7.61\%   \\
        12 & \cellcolor[HTML]{EBEBEB}15.18  & 18.51  & \cellcolor[HTML]{EBEBEB}17.99\%  & 17.53  & \multicolumn{1}{r|}{\cellcolor[HTML]{EBEBEB}13.41\%}  & 26 & \cellcolor[HTML]{EBEBEB}92.8   & 142.27 & \cellcolor[HTML]{EBEBEB}34.77\% & 103.83 & \cellcolor[HTML]{EFEFEF}10.62\%  \\
        13 & \cellcolor[HTML]{EBEBEB}18.0   & 22.62  & \cellcolor[HTML]{EBEBEB}20.42\%  & 19.55  & \multicolumn{1}{r|}{\cellcolor[HTML]{EBEBEB}7.93\%}   & 27 & \cellcolor[HTML]{EBEBEB}99.26  & 173.8  & \cellcolor[HTML]{EBEBEB}42.89\% & 118.84 & \cellcolor[HTML]{EFEFEF}16.48\%  \\
        14 & \cellcolor[HTML]{EBEBEB}20.16  & 27.33  & \cellcolor[HTML]{EBEBEB}26.23\%  & 21.99  & \multicolumn{1}{r|}{\cellcolor[HTML]{EBEBEB}8.32\%}   & 28 & \cellcolor[HTML]{EBEBEB}128.41 & 302.32 & \cellcolor[HTML]{EBEBEB}57.53\% & 132.83 & \cellcolor[HTML]{EFEFEF}3.33\%   \\
        15 & \cellcolor[HTML]{EBEBEB}24.21  & 36.74  & \cellcolor[HTML]{EBEBEB}34.10\%  & 24.56  & \multicolumn{1}{r|}{\cellcolor[HTML]{EBEBEB}1.43\%}   & 29 & \cellcolor[HTML]{EBEBEB}176.23 & 489.83 & \cellcolor[HTML]{EBEBEB}64.02\% & 453.21 & \cellcolor[HTML]{EFEFEF}61.12\%  \\ \midrule

        \multicolumn{12}{c}{\textbf{GCC-4.4.0}} \\ \midrule
        ID & \cellcolor[HTML]{EBEBEB}\proj & \tamer & \cellcolor[HTML]{EBEBEB}$\Delta_{\%}(\tamer)$ & \dthree & \multicolumn{1}{r|}{\cellcolor[HTML]{EBEBEB}$\Delta_{\%}(\dthree)$} & ID & \cellcolor[HTML]{EBEBEB}\proj & \tamer & \cellcolor[HTML]{EBEBEB}$\Delta_{\%}(\tamer)$ & \dthree & \cellcolor[HTML]{EBEBEB}$\Delta_{\%}(\dthree)$ \\ \midrule
        2  & \cellcolor[HTML]{EBEBEB}2.00  & 3.90  & \cellcolor[HTML]{EBEBEB}48.72\% & 2.72  & \multicolumn{1}{r|}{\cellcolor[HTML]{EBEBEB}26.47\%} & 7  & \cellcolor[HTML]{EBEBEB}7.00  & 14.46 & \cellcolor[HTML]{EBEBEB}51.59\% & 9.41  & \cellcolor[HTML]{EFEFEF}25.61\% \\
        3  & \cellcolor[HTML]{EBEBEB}3.00  & 5.92  & \cellcolor[HTML]{EBEBEB}49.32\% & 4.36  & \multicolumn{1}{r|}{\cellcolor[HTML]{EBEBEB}31.19\%} & 8  & \cellcolor[HTML]{EBEBEB}8.00  & 16.69 & \cellcolor[HTML]{EBEBEB}52.07\% & 10.74 & \cellcolor[HTML]{EFEFEF}25.51\% \\
        4  & \cellcolor[HTML]{EBEBEB}4.00  & 7.94  & \cellcolor[HTML]{EBEBEB}49.62\% & 5.44  & \multicolumn{1}{r|}{\cellcolor[HTML]{EBEBEB}26.47\%} & 9  & \cellcolor[HTML]{EBEBEB}9.00  & 18.43 & \cellcolor[HTML]{EBEBEB}51.17\% & 11.80 & \cellcolor[HTML]{EFEFEF}23.73\% \\
        5  & \cellcolor[HTML]{EBEBEB}5.00  & 8.97  & \cellcolor[HTML]{EBEBEB}44.26\% & 6.61  & \multicolumn{1}{r|}{\cellcolor[HTML]{EBEBEB}24.36\%} & 10 & \cellcolor[HTML]{EBEBEB}10.00 & 20.54 & \cellcolor[HTML]{EBEBEB}51.31\% & 13.17 & \cellcolor[HTML]{EFEFEF}24.07\% \\
        6  & \cellcolor[HTML]{EBEBEB}6.00  & 10.92 & \cellcolor[HTML]{EBEBEB}45.05\% & 8.11  & \multicolumn{1}{r|}{\cellcolor[HTML]{EBEBEB}26.02\%} & 11 & \cellcolor[HTML]{EBEBEB}16.00 & 24.25 & \cellcolor[HTML]{EBEBEB}34.02\% & 17.81 & \cellcolor[HTML]{EFEFEF}10.16\% \\ \midrule

        \multicolumn{12}{c}{\textbf{GCC-4.5.0}} \\ \midrule
        ID & \cellcolor[HTML]{EBEBEB}\proj & \tamer & \cellcolor[HTML]{EBEBEB}$\Delta_{\%}(\tamer)$ & \dthree & \multicolumn{1}{r|}{\cellcolor[HTML]{EBEBEB}$\Delta_{\%}(\dthree)$} & ID & \cellcolor[HTML]{EBEBEB}\proj & \tamer & \cellcolor[HTML]{EBEBEB}$\Delta_{\%}(\tamer)$ & \dthree & \cellcolor[HTML]{EBEBEB}$\Delta_{\%}(\dthree)$ \\ \midrule
        2  & \cellcolor[HTML]{EBEBEB}2.00  & 2.20  & \cellcolor[HTML]{EBEBEB}9.09\%  & 2.10  & \multicolumn{1}{r|}{\cellcolor[HTML]{EBEBEB}4.76\%}   & 5  & \cellcolor[HTML]{EBEBEB}7.14  & 8.81  & \cellcolor[HTML]{EBEBEB}18.96\% & 7.16  & \cellcolor[HTML]{EFEFEF}0.28\%  \\
        3  & \cellcolor[HTML]{EBEBEB}3.12  & 4.50  & \cellcolor[HTML]{EBEBEB}30.67\% & 3.13  & \multicolumn{1}{r|}{\cellcolor[HTML]{EBEBEB}0.32\%}   & 6  & \cellcolor[HTML]{EBEBEB}9.66  & 12.47 & \cellcolor[HTML]{EBEBEB}22.53\% & 16.02 & \cellcolor[HTML]{EFEFEF}39.70\% \\
        4  & \cellcolor[HTML]{EBEBEB}5.24  & 6.53  & \cellcolor[HTML]{EBEBEB}19.75\% & 4.80  & \multicolumn{1}{r|}{\cellcolor[HTML]{EBEBEB}-9.17\%}  & 7  & \cellcolor[HTML]{EBEBEB}11.33 & 19.54 & \cellcolor[HTML]{EBEBEB}42.02\% & 21.64 & \cellcolor[HTML]{EFEFEF}47.64\% \\ \midrule

        \multicolumn{12}{c}{\textbf{LLVM-2.8.0}} \\ \midrule
        ID & \cellcolor[HTML]{EBEBEB}\proj & \tamer & \cellcolor[HTML]{EBEBEB}$\Delta_{\%}(\tamer)$ & \dthree & \multicolumn{1}{r|}{\cellcolor[HTML]{EBEBEB}$\Delta_{\%}(\dthree)$} & ID & \cellcolor[HTML]{EBEBEB}\proj & \tamer & \cellcolor[HTML]{EBEBEB}$\Delta_{\%}(\tamer)$ & \dthree & \cellcolor[HTML]{EBEBEB}$\Delta_{\%}(\dthree)$ \\ \midrule
        2  & \cellcolor[HTML]{EBEBEB}2.02  & 2.13  & \cellcolor[HTML]{EBEBEB}5.16\%  & 2.05  & \multicolumn{1}{r|}{\cellcolor[HTML]{EBEBEB}1.46\%}  & 4  & \cellcolor[HTML]{EBEBEB}7.23  & 39.29 & \cellcolor[HTML]{EBEBEB}81.60\% & 6.09  & \cellcolor[HTML]{EFEFEF}-18.72\% \\
        3  & \cellcolor[HTML]{EBEBEB}3.54  & 4.05  & \cellcolor[HTML]{EBEBEB}12.59\% & 3.65  & \multicolumn{1}{r|}{\cellcolor[HTML]{EBEBEB}3.01\%}  & 5  & \cellcolor[HTML]{EBEBEB}10.14 & 46.66 & \cellcolor[HTML]{EBEBEB}78.27\% & 11.06 & \cellcolor[HTML]{EFEFEF}8.32\%   \\ \midrule

        \multicolumn{12}{c}{\textbf{GCC-13.1.0}} \\ \midrule
        ID & \cellcolor[HTML]{EBEBEB}\proj & \tamer & \cellcolor[HTML]{EBEBEB}$\Delta_{\%}(\tamer)$ & \dthree & \multicolumn{1}{r|}{\cellcolor[HTML]{EBEBEB}$\Delta_{\%}(\dthree)$} & ID & \cellcolor[HTML]{EBEBEB}\proj & \tamer & \cellcolor[HTML]{EBEBEB}$\Delta_{\%}(\tamer)$ & \dthree & \cellcolor[HTML]{EBEBEB}$\Delta_{\%}(\dthree)$ \\ \midrule
        2  & \cellcolor[HTML]{EBEBEB}2.00  & 2.62  & \cellcolor[HTML]{EBEBEB}23.66\% & 2.46  & \multicolumn{1}{r|}{\cellcolor[HTML]{EBEBEB}18.70\%} & 5  & \cellcolor[HTML]{EBEBEB}5.00 & 7.74  & \cellcolor[HTML]{EBEBEB}35.40\% & 6.78 & \cellcolor[HTML]{EFEFEF}26.25\% \\
        3  & \cellcolor[HTML]{EBEBEB}3.00  & 4.30  & \cellcolor[HTML]{EBEBEB}30.23\% & 3.54  & \multicolumn{1}{r|}{\cellcolor[HTML]{EBEBEB}15.25\%} & 6  & \cellcolor[HTML]{EBEBEB}7.08 & 10.54 & \cellcolor[HTML]{EBEBEB}32.83\% & 8.21 & \cellcolor[HTML]{EFEFEF}13.76\% \\
        4  & \cellcolor[HTML]{EBEBEB}4.00  & 5.82  & \cellcolor[HTML]{EBEBEB}31.27\% & 4.95  & \multicolumn{1}{r|}{\cellcolor[HTML]{EBEBEB}19.19\%} & 7  & \cellcolor[HTML]{EBEBEB}9.26 & 19.86 & \cellcolor[HTML]{EBEBEB}53.37\% & 9.98 & \cellcolor[HTML]{EFEFEF}7.21\%  \\ \bottomrule
        \end{tabular}%
    }
\end{table*}

The findings from RQ1 and RQ2 give rise to \proj,
a new bisection-based approach to compiler bug deduplication,
that follows a fundamentally different technical
direction from existing analysis-based techniques.
\proj leverages the failure-inducing commits obtained from bisection
as the primary criterion of deduplication, and identifies the specific
bug-triggering optimizations as a complementary
factor to mitigate false negatives.
Given the encouraging finding from RQ1 and RQ2, we stand poised to
evaluate \proj on a larger scale, across all five datasets introduced
earlier, to further validate its effectiveness and practical utility.
In this comprehensive evaluation, we employ a more fine-grained metric,
namely \emph{wasted effort}.
Compared to the visually intuitive bug discovery curve,
the \emph{wasted effort} metric provides a quantitative measure,
indicating precisely how many test cases must be examined before
each distinct bug is identified. A lower wasted effort reflects
a more effective deduplication process, directly highlighting the
reduction in human debugging effort achieved by the deduplication technique.

We compare \proj with \tamer and \dthree, and present the detailed
results in \cref{tab:waste-effort}.
The "ID" column represents each distinct bug identified, whereas,
it worth a note that an "ID" does not necessarily correspond to the same
bug across different techniques, as the order of bug identification
may vary. The columns $\Delta_{\%}(\tamer)$ and $\Delta_{\%}(\dthree)$
indicate the relative improvement of \proj over \tamer and \dthree, respectively.
Overall, the results clearly demonstrate that \proj significantly
outperforms both \tamer and \dthree by saving the wasted effort in
identifying each distinct bug.
Specifically, \proj outperforms \tamer in \numBugsBetterThanTamer out of
\numAllBuugsInTable cases by reducing an average of \saveEffortAvgTamer
(\saveEffortAvgPercentageTamer) test case examinations to identify each bug.
Similarly, \proj achieves better performance than \dthree
in \numBugsBetterThanDthree cases, saving an average of
\saveEffortAvgDthree (\saveEffortAvgPercentageDthree)
test case examinations per bug.
While a few cases exist where \tamer and \dthree achieve slightly
better results, the differences remain marginal.
In fact, \tamer and \dthree only save an average of \tamerBetterCasesSaveEffortAvg
and \dthreeBetterCasesSaveEffortAvg wasted effort than \proj in
these cases, respectively.

Furthermore, to validate the statistical significance of the results,
we conduct pairwise Wilcoxon signed-rank tests~\cite{woolson2005wilcoxon}
on the wasted effort data from the two larger datasets (\gcc-4.3.0 and
\gcc-4.4.0); the other two datasets have too few unique bugs to justify
statistical analysis.
The p-values are all less than 0.05
(\tamerPValueGccFourThreeZero and \dthreePValueGccFourThreeZero
for \tamer and \dthree on \gcc-4.3.0, respectively, and \tamerPValueGccFourFourZero
for both \tamer and \dthree on \gcc-4.4.0),
indicating that the improvements of \proj over both \tamer and
\dthree are statistically significant.

\section{Generality}
\label{sec:generality}
The results presented in \cref{sec:effectiveness} successfully demonstrate
the effectiveness of bisection-based bug deduplication.
However, effectiveness alone does not ensure the practical utility of
a bug deduplication approach; other properties, particularly generality,
also play a critical role.
As discussed in \cref{subsec:existing-techniques},
analysis-based techniques often face criticism regarding their
generality due to two primary issues:
(1) domain-specificity, requiring additional human
    effort to adapt them to different domains, and
(2) ineffective or even infeasible on unminimized test programs,
    relying on test input minimization as a pre-processing step.
Bisection-based bug deduplication inherently addresses the first issue
because it can be directly adapted to different domains without extra
effort, provided the target compiler is maintained in a version
control system and supports multiple configuration modes.
However, the second issue remains concerning.
While bisection theoretically should not be affected by the
minimization status of test programs, practical scenarios may differ.
In particular, irrelevant fragments in unminimized programs might
lead to variations in compiler behavior across different versions,
resulting in unexpected situations during the bisection process.
For instance, we encountered a scenario where a program
triggering a miscompilation bug caused intermediate compiler versions
to crash during bisection.
Such crashes prevent us from determining whether the miscompilation
bug is present in the current version, as they may result from an
earlier-stage failure triggered by code irrelevant to the bug
of interest.
These unexpected situations could increase false positives, wherein
identical bugs are incorrectly identified as distinct due to
inconsistent bisection outcomes, ultimately reducing deduplication effectiveness.
To evaluate the generality of bisection-based bug deduplication,
we investigate the following research question:
\begin{itemize}[leftmargin=*]
    \item \textbf{RQ3:} \rqThree
\end{itemize}

\begin{figure*}[h!]
    \centering
    \begin{subfigure}[b]{0.325\columnwidth}
        \includegraphics[width=\columnwidth]{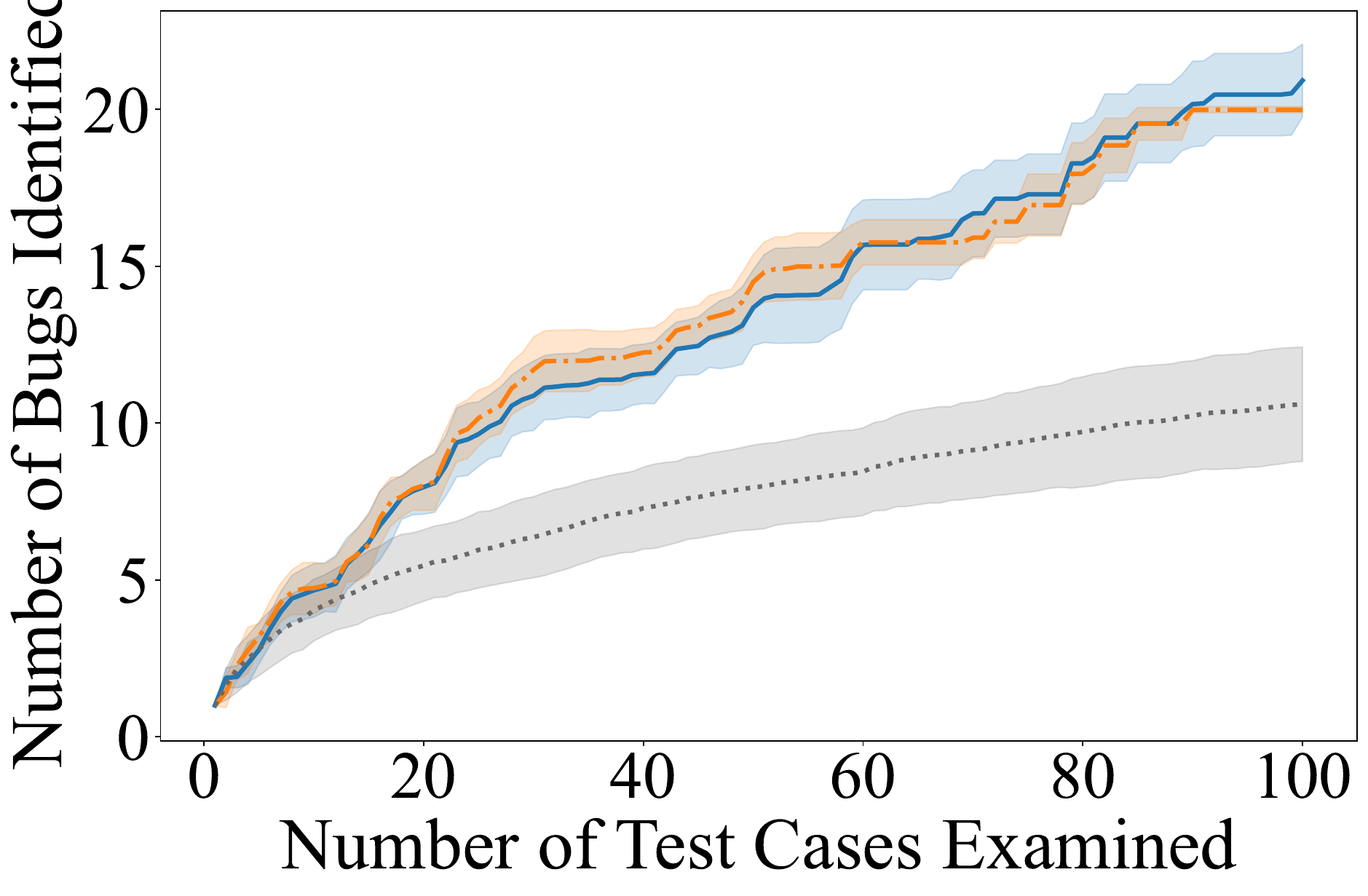}
        \caption{
       	    First 100 tests on \gcc-4.3.0
       	}
       \label{subfig:rq1-gcc430-orig-curve-first-100}
    \end{subfigure}
        \hfil
    \begin{subfigure}[b]{0.325\columnwidth}
        \includegraphics[width=\columnwidth]{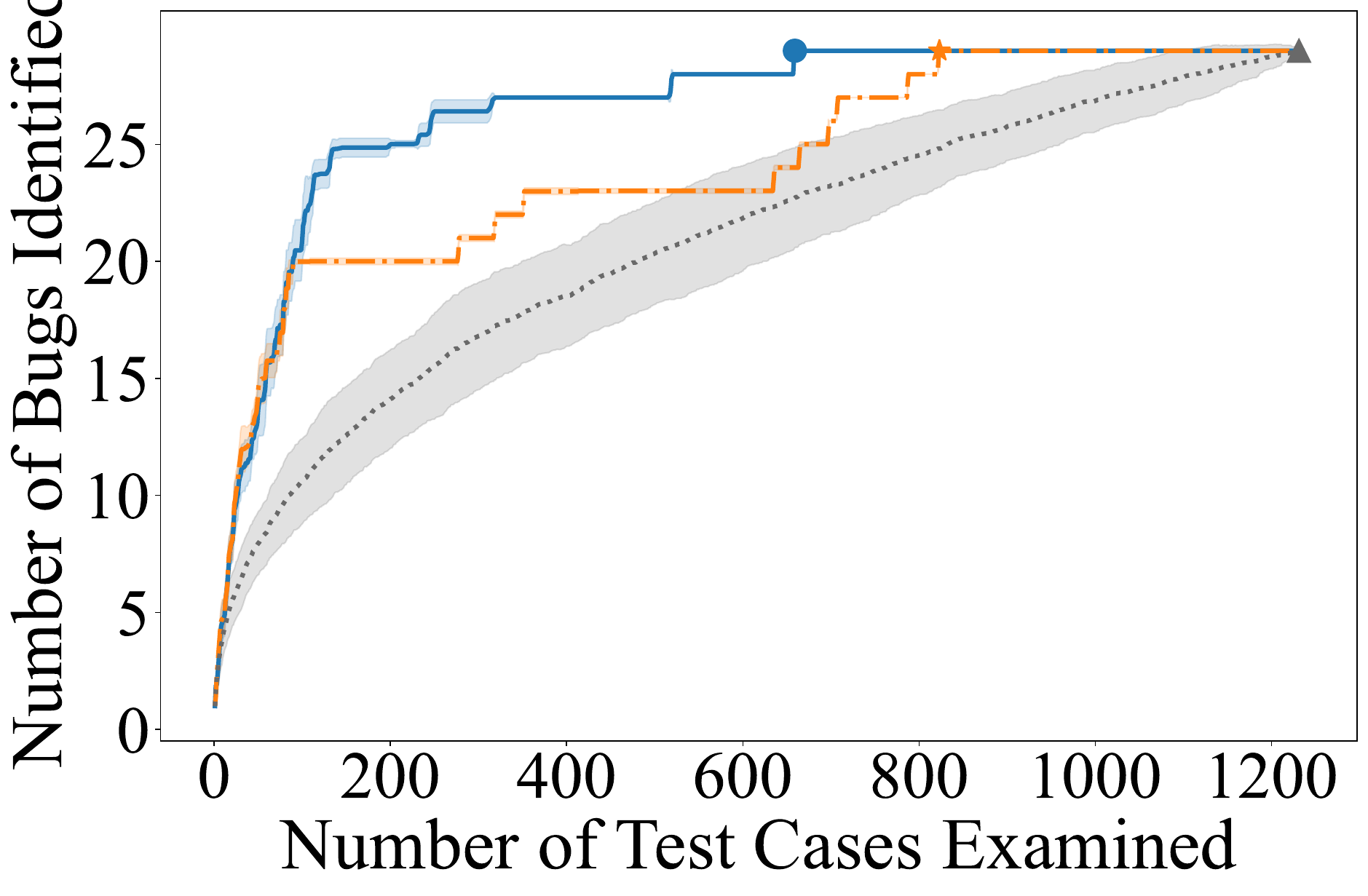}
        \caption{
            All tests on \gcc-4.3.0
        }
        \label{subfig:rq1-gcc430-orig-curve-all-tests}
    \end{subfigure}
        \hfil
    \begin{subfigure}[b]{0.325\columnwidth}
        \includegraphics[width=\columnwidth]{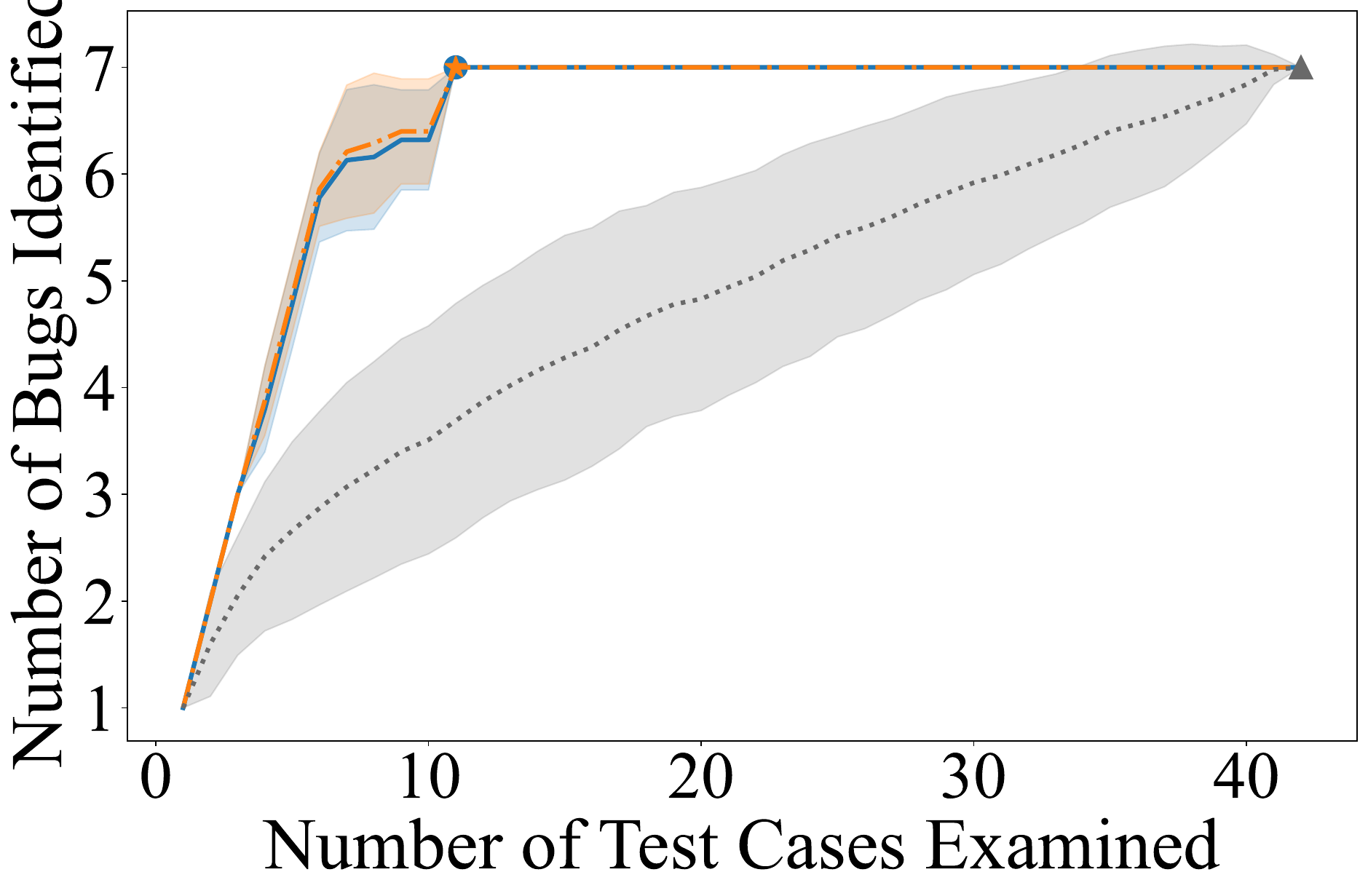}
        \caption{
            \small{
                All tests on \gcc-13.1.0
            }
        }
        \label{subfig:rq1-gcc1310-orig-curve-all-tests}
    \end{subfigure}

    \caption{
        Bug discovery curves of \bisectionSole, \proj,
        and the baseline on the unminimized \gcc-4.3.0 and \gcc-13.1.0 datasets.
    }
    \label{fig:rq3-bug-discovery-curve}

\end{figure*}

To answer this RQ, we perform bisection-based bug deduplication
(both \bisectionSole and \proj)
on the original unminimized test programs in the \gcc-4.3.0 and \gcc-13.1.0 datasets,
which are the only datasets for which the original unminimized test programs are available.

As discussed in \cref{subsec:existing-techniques}, \dthree is
practically unrealistic to apply to unminimized programs.
Additionally, the authors of \tamer also acknowledged that they failed
to conduct experiments on unminimized test cases due to the
excessively large data volume involved~\cite{tamer}. Therefore,
we limit our comparison to a random selection strategy for unminimized
test programs. The resulting bug discovery curves are presented in
\cref{fig:rq3-bug-discovery-curve}, clearly illustrating that
bisection-based bug deduplication remains effective even
with unminimized test programs.
For example, on dataset \gcc-4.3.0,
\proj successfully identifies
21 bugs during the first 100 test programs,
while random selection can only discover 10 bugs.
Furthermore, \proj
requires \bisectionWithOptFindAllBugsOrig test case examinations
to identify all the \benchmarkGCCFourThreeZeroBugs bugs, saving
\bisectionWithOptSaveEffortThanRandomOrig of the effort compared
to random selection.
Notably, exploiting the bug-triggering optimizations can still
effectively improve the overall performance by reducing false negatives.
However, as anticipated, irrelevant fragments in
unminimized programs negatively impacts deduplication performance,
primarily due to increased false positives.
In response to this challenge, we
provide practical suggestions in \cref{subsec:suggestions} to
help balance deduplication effectiveness with the overhead of minimization.
On the more recent \gcc-13.1.0 dataset,
the bisection-based approach remains highly effective,
identifying all \benchmarkGCCThirdteenBugs bugs with an average of
only 9.75 test case examinations,
and reducing effort by 71.17\% compared with random selection.
Ultimately, bisection-based bug deduplication offers a valuable
alternative to analysis-based techniques, particularly in scenarios
where effective test input minimization is unavailable.

\myFinding{
    Bisection-based bug
    deduplication remains effective even without test input
    minimization, demonstrating superior generality.
    Nevertheless, irrelevant code fragments in original
    test programs can complicate the bisection process,
    thereby reducing deduplication performance.
    The trade-off between deduplication effectiveness and
    minimization overhead should be carefully balanced in practice.
}

\section{Efficiency}
\label{sec:efficiency}
\begin{figure*}[h!]
    \centering
    \begin{subfigure}[b]{0.325\columnwidth}
        \includegraphics[width=\columnwidth]{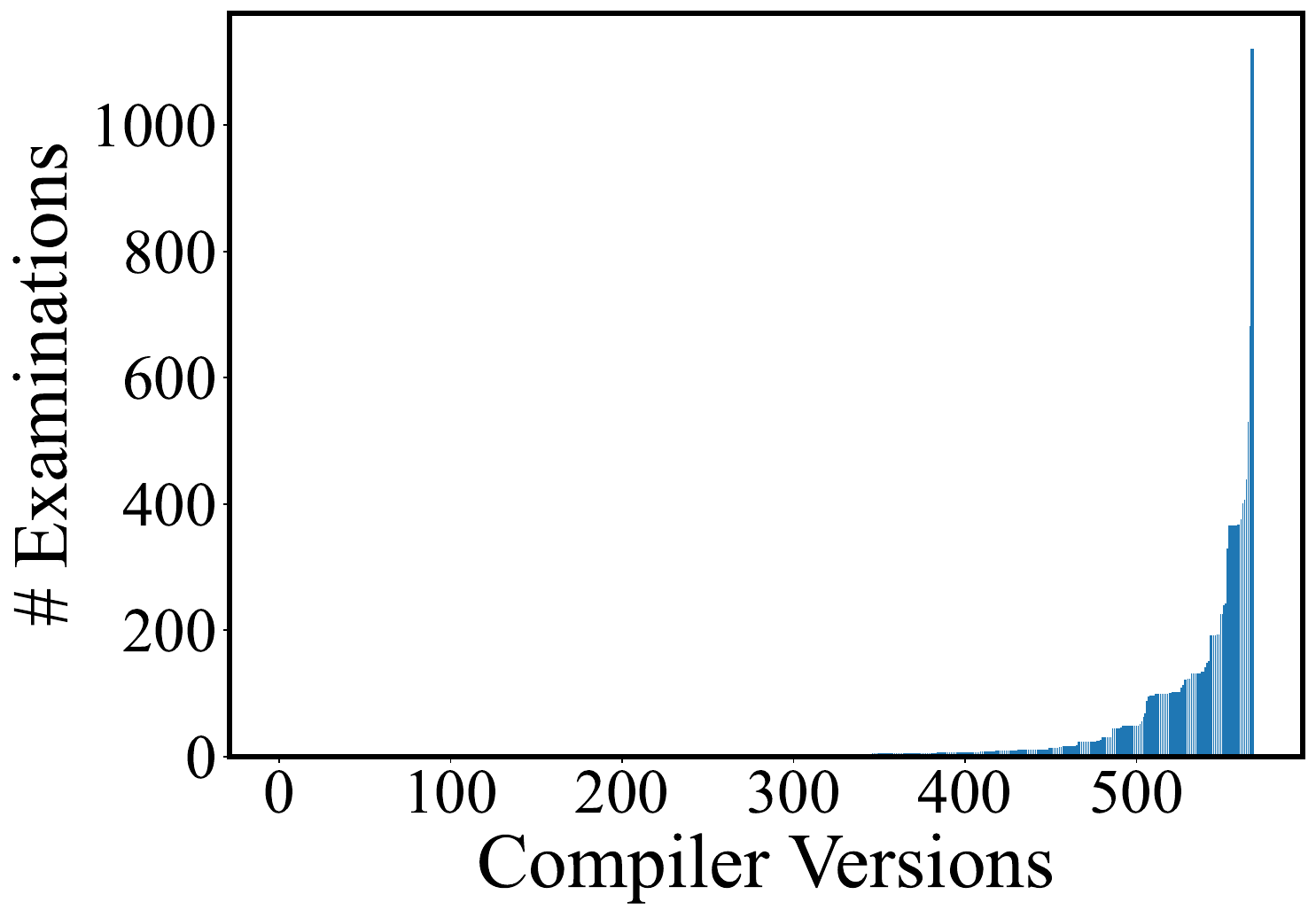}
       \caption{
       	    GCC-4.3.0
       	}
       \label{subfig:rq4-gcc430-bar}
    \end{subfigure}
    \hfil
    \begin{subfigure}[b]{0.325\columnwidth}
        \includegraphics[width=\columnwidth]{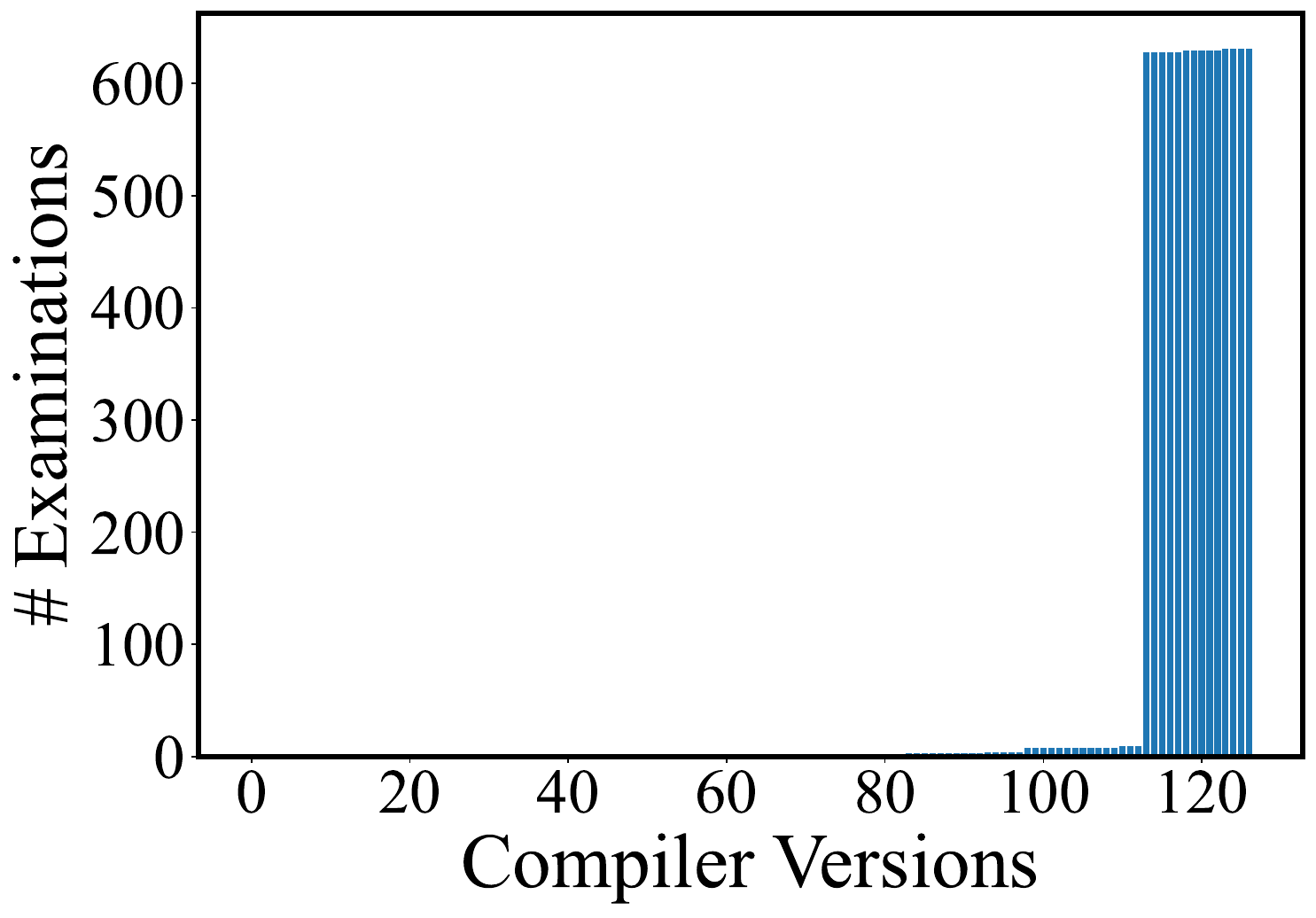}
        \caption{
            GCC-4.4.0
        }
        \label{subfig:rq4-gcc440-bar}
    \end{subfigure}

    \begin{subfigure}[b]{0.325\columnwidth}
        \includegraphics[width=\columnwidth]{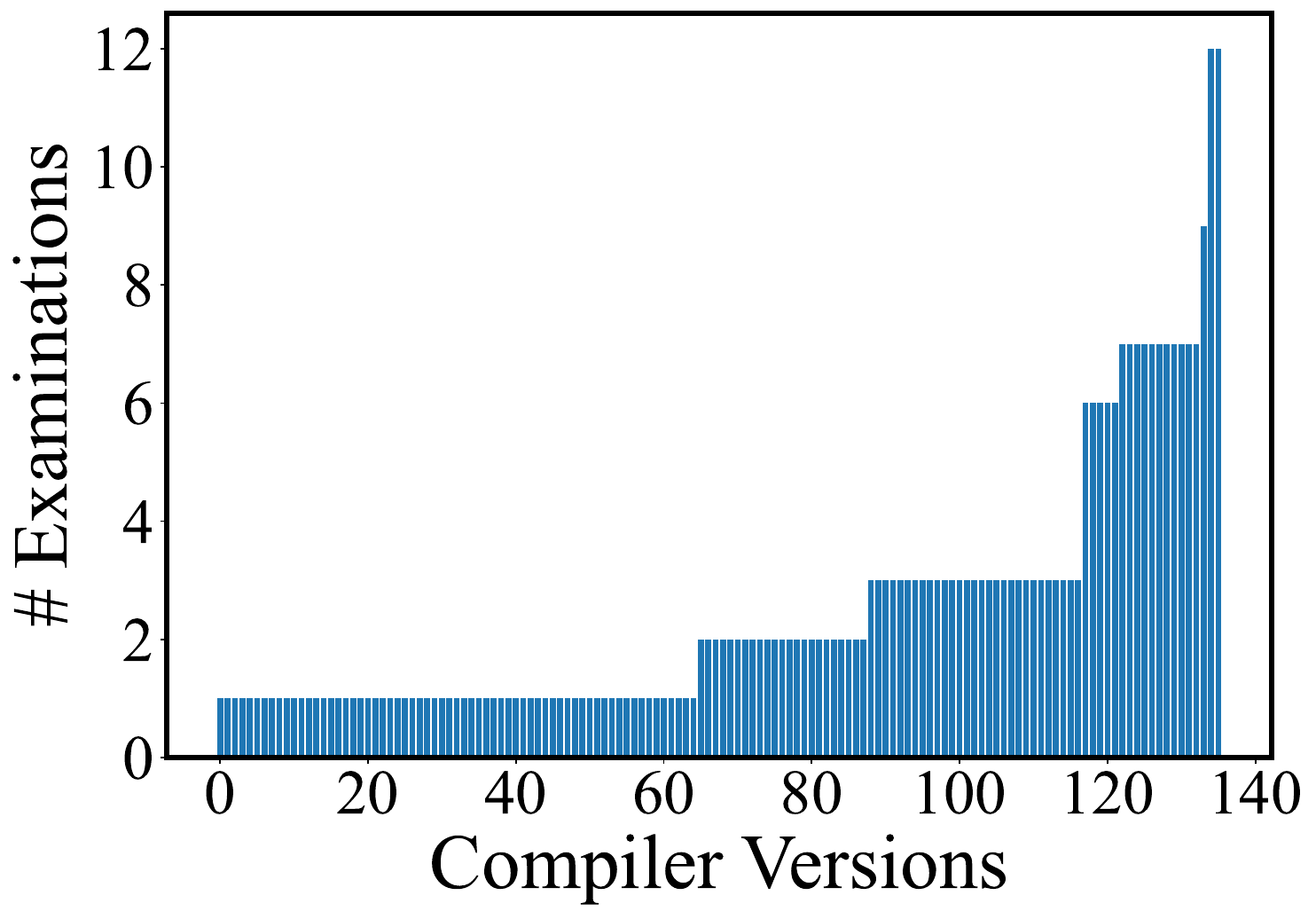}
        \caption{
            GCC-4.5.0
        }
        \label{subfig:rq4-gcc450-bar}
    \end{subfigure}
    \hfil
    \begin{subfigure}[b]{0.325\columnwidth}
        \includegraphics[width=\columnwidth]{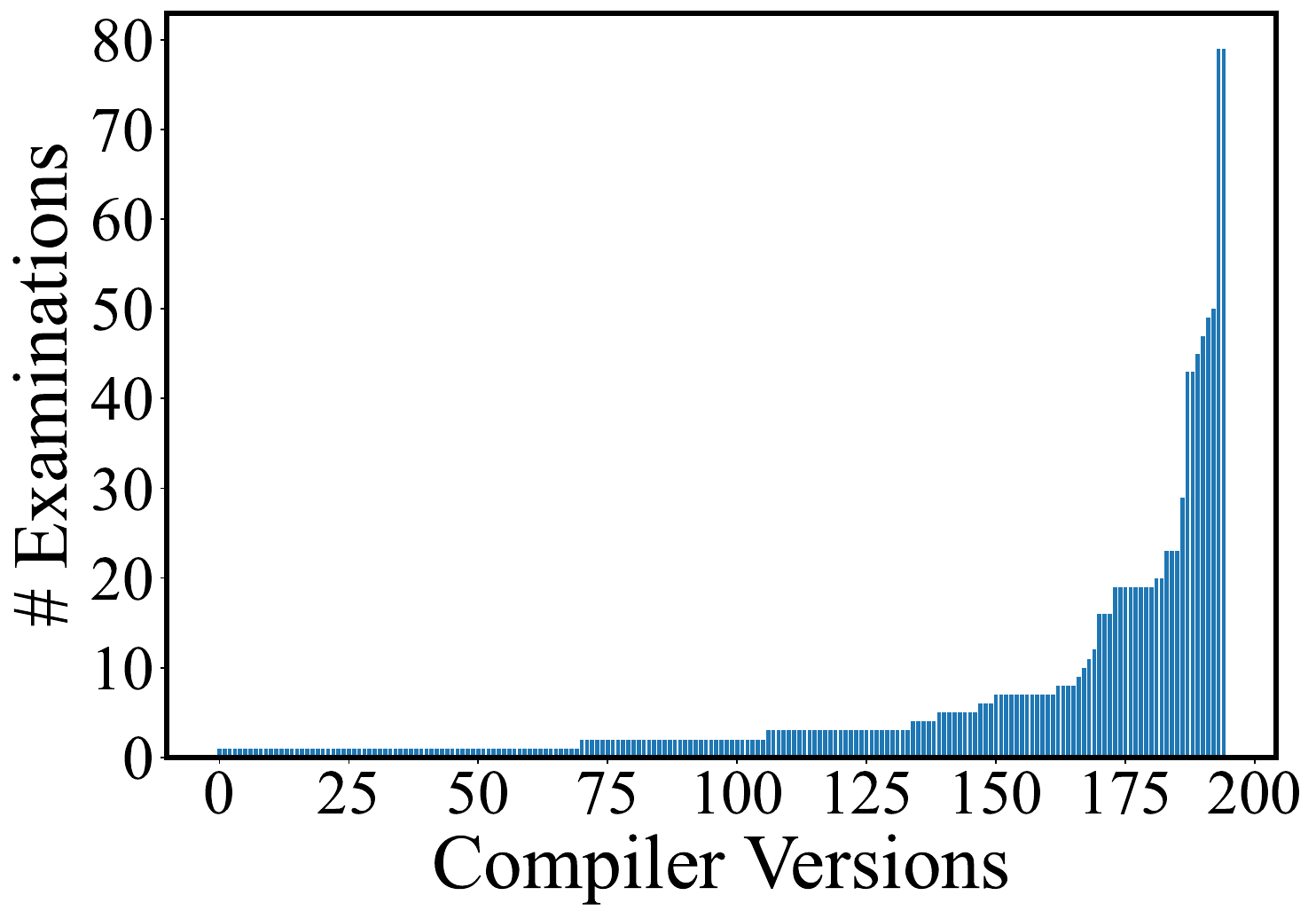}
        \caption{
            LLVM-2.8.0
        }
        \label{subfig:rq4-llvm280-bar}
    \end{subfigure}
    \hfil
    \begin{subfigure}[b]{0.325\columnwidth}
        \includegraphics[width=\columnwidth]{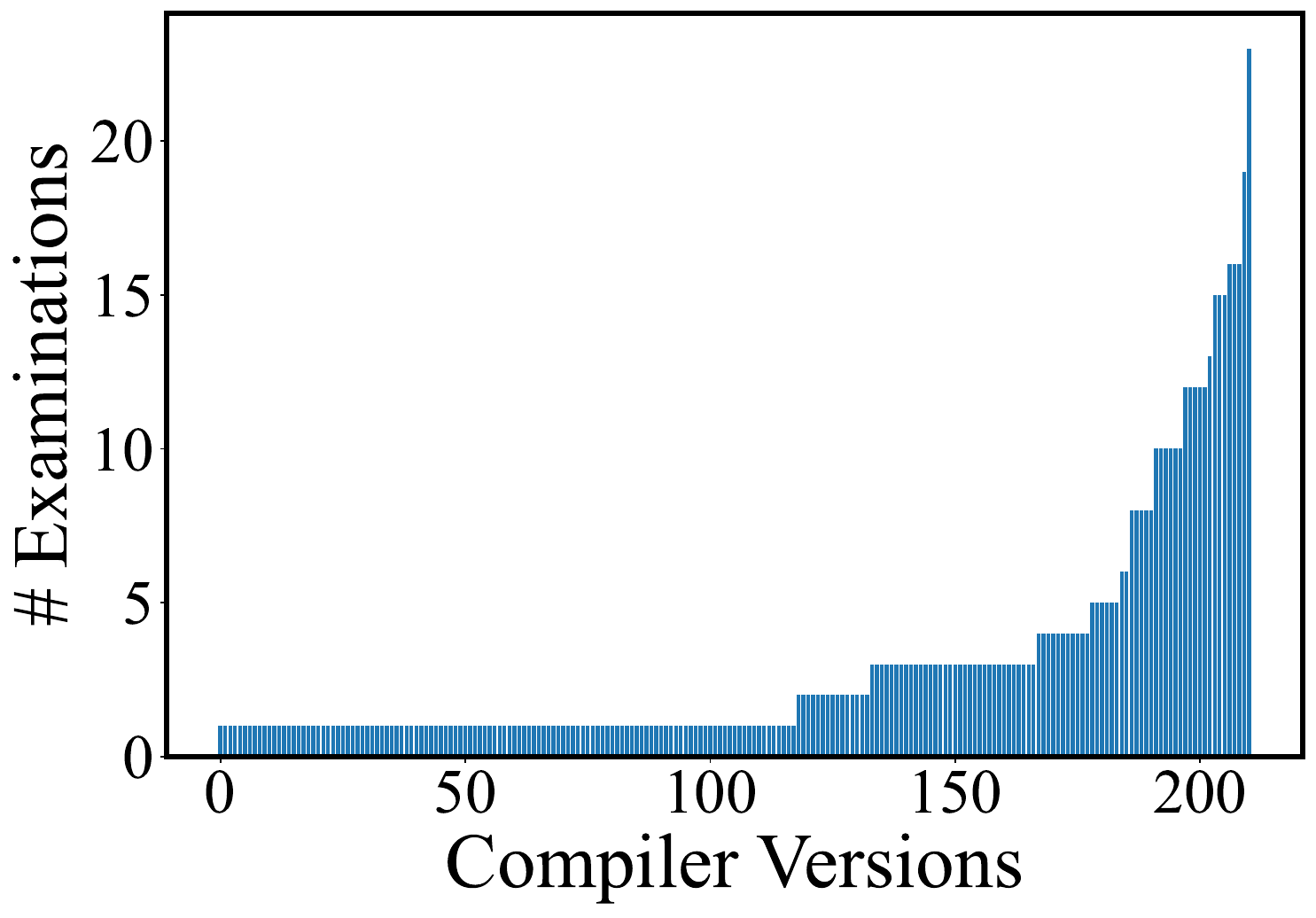}
        \caption{
            GCC-13.1.0
        }
        \label{subfig:rq4-gcc1310-bar}
    \end{subfigure}
    \caption{
        Number of times each compiler version being examined by
        different test programs during bisection in each datasets.
    }
    \label{fig:rq4-commit-hash-bar}
\end{figure*}

Another important consideration when using bisection to deduplicate
bugs is its efficiency.
Bisection requires building a series of intermediate compiler versions
from the commit history, which constitutes the most time-consuming
component of the entire deduplication process.
For a dataset containing $n$ test programs, and an average bisection
range of $m$ commits, the theoretical maximum number of compiler
versions built would be $n \cdot \log_{2}m$. Given the typically
large volume of test cases, efficiency becomes a crucial factor
influencing the practical utility of bug deduplication through bisection.
Our last research question aims to investigate this issue:
\begin{itemize}[leftmargin=*]
    \item \textbf{RQ4:} \rqFour
\end{itemize}

Building a compiler from source code can take varying amounts of time
on different machines, depending on the hardware configuration and the
number of available cores. To formalize the efficiency evaluation,
we treat the time required for a single compiler build as a constant,
and measure efficiency by counting how many compiler versions must be
built during the bisection process.
Notably, although the same compiler version may be examined multiple
times by different programs, it only needs to be built once and
can then be cached for reuse---a strategy adopted in our implementation.
Given the large number of test programs in our datasets, this
caching mechanism significantly reduces the total number of builds
required, thereby enhancing efficiency.

\begin{table}[]
    \centering
    \caption{Number of compiler versions built during the whole bisection-based bug deduplication process.}
    \label{tab:efficiency}
    \setlength{\tabcolsep}{7pt}
    \resizebox{0.87\columnwidth}{!}{%
    \begin{tabular}{rccccc}
    \toprule
                        & \textbf{GCC-4.3.0} & \textbf{GCC-4.4.0} & \textbf{GCC-4.5.0} & \textbf{LLVM-2.8.0} & \textbf{GCC-13.1.0} \\ \midrule
    \textit{\quad\textbf{\#} Versions}         & 569       & 127       & 136       & 195         & 211\quad        \\
    \textit{\quad\textbf{\#} Versions per Case} & 0.46      & 0.19      & 5.23      & 2.44        & 5.02\quad       \\ \bottomrule
    \end{tabular}%
    }

\end{table}

The number of compiler versions built during the entire deduplication
process for each dataset is presented in \cref{tab:efficiency}. Among
the five datasets, the average number of compiler versions
built per case ranges from 0.19 to 5.23, typically completing
with a relatively short timeframe.
Therefore, although our approach theoretically has high worst-case
complexity, it remains practically efficient.
\cref{fig:rq4-commit-hash-bar}
illustrates how frequently each compiler version is examined by
various test programs during bisection,
where the x-axis represents the number of compiler versions being
examined, and the y-axis represents the number of times a
version is examined by different test programs.
The results show that most compiler versions are examined
multiple times by different test programs.
The data confirms that a
single compiler build can effectively serve multiple test programs,
aligning with our expectations.
In practice, the higher the
proportion of duplicated bugs, the fewer compiler versions need
to be built, as programs triggering the same bug generally
produce similar bisection traces.
Moreover, the efficiency can be further improved by maintaining
a list of pre-built compiler binaries for quickly narrowing down
the bisection range.
Leveraging incremental build~\cite{randrianaina2022benefits} can also
bring efficiency benefits by accelerating each build process.

\myFinding{
    The average number of compiler versions need to be built
    during the deduplication process with bisection is
    relatively small, ranging from 0.19 to 5.23 per test program.
    This indicates that bisection-based bug deduplication
    maintains high practical efficiency.
    The most time-consuming aspect---building multiple compiler
    versions---is well mitigated by reusing the same version across
    multiple test programs.
}

\section{Discussion}
\label{sec:discussion}

\myTightParagraph{Practical Suggestions}
\label{subsec:suggestions}
Our study confirms the feasibility of using bisection to deduplicate
compiler bugs in terms of effectiveness, generality, and efficiency.
Prior research has typically treated bug deduplication and test input
minimization as separate tasks, relying heavily on test input
minimizers to enable effective deduplication,
thus inevitably suffering from the overhead of minimization.
In contrast, our approach provides
an alternative solution that reduces this dependency.
Nevertheless, noise present in unminimized test programs can still
interfere with the bisection process, potentially diminishing
deduplication effectiveness.
To address this,
we propose the following suggestions for practitioners.

Rather than minimizing the entire set of bug-triggering test programs
upfront, we recommend first applying bisection-based
deduplication directly on the original test programs.
Once preliminary deduplication results are obtained,
practitioners can select a smaller subset of test programs
for inclusion in bug reports---only this subset requires minimization.
The number of selected programs can be adjusted
according to the development team's available resources.
After the reported bugs are fixed, remaining duplicate test
programs can be efficiently filtered by executing them on the
patched compiler version. This workflow can be iteratively applied
until all bugs are addressed.
Crucially, the deduplication algorithm remains central to this process.
Without it, randomly selecting test programs for reporting may result
in redundant submissions of the same bug---wasting valuable developer
time on diagnosing duplicate reports.
An effective deduplication system ensures that the selected
bugs for reporting are as diverse as possible, thereby maximizing
developer productivity and minimizing duplicated effort.

\myTightParagraph{Originality \& Contributions}
This study successfully demonstrates
the feasibility and effectiveness of using bisection
for compiler bug deduplication.
Our findings show that leveraging bisection to
identify failure-inducing commits provides
a useful signal for bug deduplication,
albeit one that requires supplementary techniques
for more accurate identification.
Admittedly, the underlying technique itself is not new.
For example,
Theodoridis \etal~\cite{theodoridis2022finding,theodoridis2024refined}
used a bisection-based approach to identify bug-inducing commits
that lead to compiler performance degradations,
and Yang \etal~\cite{yang2025using} revisited automated compiler
fault isolation and showed that bisection can effectively serve as
a lightweight alternative to more sophisticated techniques.
Nevertheless, the originality of our work lies in being the
\emph{first systematic study to investigate bisection as a signal for
compiler bug deduplication and to directly compare it empirically
with prior deduplication techniques}.
This constitutes a novel application of the technique.
While prior work has proposed sophisticated and often complex algorithms
for this task, our study shows that this long-standing practical challenge
can be addressed effectively with a lightweight and straightforward approach.
In this sense, our study
\emph{
    highlights the value of re-evaluating simple
    techniques before resorting to more elaborate solutions
}.
We believe this \emph{simplicity is a strength},
and our findings suggest that such an approach can
serve as a strong and practical baseline
for future work on compiler bug deduplication.

\myTightParagraph{Generality}
A primary threat to the generality of bisection-based bug deduplication
is that it requires bugs to be regressions, \ie,
there must exist both a \emph{good} version and a \emph{bad} version.
As a result, our approach is not applicable to non-regression bugs,
such as bugs that persist throughout the entire history of the compiler.
Nevertheless, we argue that the proposed approach remains highly
applicable in the context of modern compilers,
where most correctness bugs are regressions.
Taking \gcc as an example, developers often indicate regression information
in bug report titles~\cite{gcc_bugzilla_wrong_code_fixed}.
Based on this convention, we found that most miscompilation bugs in \gcc
are regressions, as reflected by titles beginning with ``\emph{xx Regression}''.
Furthermore, because our approach mainly targets the deduplication of
compiler testing results, we manually examined the bugs reported in
several major compiler testing
papers~\cite{sun2016toward,emi,sun2016finding,li2024boosting,ni2025legofuzz}.
These papers typically report the ``ages'' of the bugs they uncover.
Although some techniques can discover long-lived bugs, \ie,
bugs that also manifest in older compiler versions,
none report bugs that persist throughout the compiler's entire history,
that is, non-regression bugs.
Based on our experience with modern compiler testing,
finding non-regression bugs in modern compilers is theoretically possible
but practically unlikely, given their long and stable development history.
Therefore, although our approach is conceptually limited to regression bugs,
it remains highly applicable in practice.

\myTightParagraph{Limitations and Future Work}
One of the limitations of our proposed approach is that it requires the
developing history of the target software available. This issue is
practically alleviated by the prevalence of version control systems
like Git. Even for close source compilers, \eg, Intel C++ Compiler~\cite{icc},
they normally maintain a private git-like repository for development,
which can be used for bisection-based bug deduplication.
To this end, the bisection-based bug deduplication approach is
practically applicable to a wide range of domains.
Another issue is the existence of false positives in the deduplication
results, where test programs associated with the same bug are
incorrectly identified as distinct.
False positives are not unique to our approach,
but rather prevalent in other bug deduplication approaches.
Future work can focus on further improving the deduplication
effectiveness by mitigating the happening of false positives.

\section{Threats to Validity}
\label{sec:threats}
\myTightParagraph{Internal Threats}
The primary internal threats stem from the correctness of the
implementation of the bisection-based approach and its evaluation process.
Our implementation leverages the well-established \mycode{git bisect}
tool, which is widely adopted in practice, thereby significantly
mitigating risks associated with the correctness of the bisection process.
Furthermore, we incorporate an additional verification step after executing
\mycode{git bisect}, explicitly ensuring that the identified commit
indeed triggers the bug and that the preceding commit remains unaffected. 
To further ensure correctness, all components of the implementation
have undergone rigorous reviews by the authors, who carefully verified
both the code and the associated experimental results.

\myTightParagraph{External Threats}
The main external threat concerns the generalizability of the results.
While bisection-based bug deduplication is conceptually applicable
across various compilers, our evaluation focuses exclusively on the
\gcc and \llvm compilers. All existing work suffers from the same issue
due to the lack of diverse datasets.
However, considering the complexity and active development of \gcc and
\llvm, we argue that our findings possess a reasonable degree of generalizability.
Specifically, the effectiveness demonstrated in these challenging
scenarios suggests that our approach may perform comparably well in other contexts.
Future evaluations extending to additional domains would be valuable
to more conclusively affirm the general applicability of our approach.

\myTightParagraph{Construct Threats}
The primary construct threat in this study stems from the datasets used.
First, the correctness of the ground truth---\ie,
the actual bug triggered by each test program---is
critical for evaluating deduplication effectiveness.
All test programs in our datasets include ground truth labels
provided by the original authors.
To enhance the reliability of these labels,
we have manually verified their correctness to the best of our ability.
Another potential threat arises from the age of the datasets.
The existing four datasets can be traced back to decades ago,
which may raise concerns about their relevance to the current
state of compiler infrastructures.
To mitigate this threat, we have constructed a new dataset
on \gcc-13.1.0, a more recent compiler version,
and included it in our evaluation.
The results from this dataset are consistent with those from the older datasets,
suggesting that the findings may still be applicable to modern compilers.

\section{Related Work}
\label{sec:related_work}
We discuss two lines of related work.

\myTightParagraph{Bug Deduplication}
Our study focuses on deduplicating compiler bugs,
making the most relevant related work \tamer~\cite{tamer},
\dthree~\cite{d3} and \transformer~\cite{transformer},
which are introduced in \cref{subsec:existing-techniques}.
Additionally, bug deduplication techniques for general-propose
software have been studied in the literature.
These approaches typically rely on analyzing the textual
descriptions of bugs---such as bug reports and error
messages---to identify duplicates.
For example, BM25F~\cite{perez2010using} is a widely used similarity
metric for detecting duplicate bug reports.
Building on this, Sun \etal~\cite{sun2011towards}
proposed REP, which incorporates both textual content and
non-textual metadata (\eg, product, component, version)
to improve similarity measurement.
They further extended this line of work by introducing
discriminative models for more accurate duplicate
detection~\cite{sun2010discriminative}.
Nguyen \etal~\cite{nguyen2012duplicate} proposed DMTM,
which combines topic-based and textual features to characterize
bug reports for deduplication.
Zou \etal~\cite{zou2016automated} introduced a multi-factor analysis
approach that integrates topic modeling, enhanced N-gram similarity,
and contextual metadata for improved accuracy.
Lin \etal~\cite{lin2016enhancements} proposed manifold correlation
features and an enhanced SVM model that significantly boosts
duplicate detection performance.
Wang \etal~\cite{wang2008approach}
presented a hybrid technique that leverages both natural
language descriptions and execution traces,
while Lerch and Mezini~\cite{lerch2013finding}
developed a proactive approach that predicts potential
duplicates using call stack structures and machine learning.
Although these techniques are effective in their respective domains,
they are not directly applicable to compiler bug deduplication.
Compiler bug reports typically contain minimal textual
information---often limited to the bug-triggering test program
and the compiler configuration used---rendering text-based
techniques ineffective. In this context, our study introduces a
novel perspective on compiler bug deduplication by proposing a
lightweight, analysis-free approach that does not rely on rich
textual content or extensive program analysis.

\myTightParagraph{Failure-inducing Commit Localization}
Localizing the exact commit that causes a test failure
is a common practice in software debugging and fixing,
and has been extensively studied in the literature.
One of the most well-known techniques is the SZZ
algorithm~\cite{sliwerski2005changes},
along with its numerous variants~\cite{tang2023neural,rosa2023comprehensive,rani2024refining,tang2025llm4szz}.
In general, given a bug-fixing commit,
the SZZ algorithm traces the version history to identify
commits that last modified the lines removed or changed
in the fix---these are considered the bug-inducing commits.
However, SZZ and its variants are not applicable in our context,
as they rely on the existence of a bug-fixing commit,
which is not available at the time of bug detection.
Other approaches have also been proposed:
Wen \etal~\cite{wen2016locus} leverage information retrieval
techniques to match bug reports,
while An \etal~\cite{an2021reducing, an2023fonte} incorporate
code coverage analysis to aid in localizing bug-inducing commits.
More recently, Tang \etal~\cite{tang2024enhancing} introduced
a semantic analysis approach that uses program slicing and semantic
reasoning to detect suspicious commits based on data flow
and control flow differences.
While effective, these techniques either depend heavily on
the quality of bug reports or involve complex program analysis,
making them unsuitable for our setting.
To this end, we adopt bisection to localize failure-inducing commits
in this study,
aligning with our goal of developing a lightweight and
analysis-free bug deduplication approach.

\section{Conclusion}
\label{sec:conclusion}

Debugging compiler bugs identified through random testing is
a painful process,
suffering from the bug deduplication problem, \ie, numerous
duplicate test cases expose the same underlying bug. Existing
deduplication algorithms are primarily analysis-based, leading to
high complexity and limited generalizability in practice.
In this paper, we conduct the first systematic study on
leveraging bisection for deduplicating compiler bugs.
Motivated by the insights gained from this study,
we propose \proj, a simple, generalizable,
and effective approach for compiler bug deduplication.
Our method outperforms the state-of-the-art analysis-based
techniques, by significantly reducing the human effort required
for debugging. We emphasize the practical value and applicability
of bisection-based bug deduplication in real-world scenarios.

\section{Data Availability}
The artifact supporting this study is publicly available at:
\href{https://anonymous.4open.science/r/BugLens-Artifact/}{https://anonymous.4open.science/r/BugLens-Artifact/}.

\newpage

\bibliographystyle{ACM-Reference-Format}
\bibliography{acmart}

@inproceedings{tamer,
  title={Taming compiler fuzzers},
  author={Chen, Yang and Groce, Alex and Zhang, Chaoqiang and Wong, Weng-Keen and Fern, Xiaoli and Eide, Eric and Regehr, John},
  booktitle={Proceedings of the 34th ACM SIGPLAN conference on Programming language design and implementation},
  pages={197--208},
  year={2013}
}

@inproceedings{transformer,
  title={Test-case reduction and deduplication almost for free with transformation-based compiler testing},
  author={Donaldson, Alastair F and Thomson, Paul and Teliman, Vasyl and Milizia, Stefano and Maselco, Andr{\'e} Perez and Karpi{\'n}ski, Antoni},
  booktitle={Proceedings of the 42nd ACM SIGPLAN International Conference on Programming Language Design and Implementation},
  pages={1017--1032},
  year={2021}
}

@inproceedings{d3,
  title={Silent compiler bug de-duplication via three-dimensional analysis},
  author={Yang, Chen and Chen, Junjie and Fan, Xingyu and Jiang, Jiajun and Sun, Jun},
  booktitle={Proceedings of the 32nd ACM SIGSOFT International Symposium on Software Testing and Analysis},
  pages={677--689},
  year={2023}
}

@article{emi,
  title={Compiler validation via equivalence modulo inputs},
  author={Le, Vu and Afshari, Mehrdad and Su, Zhendong},
  journal={ACM Sigplan Notices},
  volume={49},
  number={6},
  pages={216--226},
  year={2014},
  publisher={ACM New York, NY, USA}
}

@inproceedings{sun2016toward,
  title={Toward understanding compiler bugs in GCC and LLVM},
  author={Sun, Chengnian and Le, Vu and Zhang, Qirun and Su, Zhendong},
  booktitle={Proceedings of the 25th international symposium on software testing and analysis},
  pages={294--305},
  year={2016}
}

@inproceedings{sun2016finding,
  title={Finding compiler bugs via live code mutation},
  author={Sun, Chengnian and Le, Vu and Su, Zhendong},
  booktitle={Proceedings of the 2016 ACM SIGPLAN international conference on object-oriented programming, systems, languages, and applications},
  pages={849--863},
  year={2016}
}

@inproceedings{csmith,
  title={Finding and understanding bugs in C compilers},
  author={Yang, Xuejun and Chen, Yang and Eide, Eric and Regehr, John},
  booktitle={Proceedings of the 32nd ACM SIGPLAN conference on Programming language design and implementation},
  pages={283--294},
  year={2011}
}

@inproceedings{creduce,
  title={Test-case reduction for C compiler bugs},
  author={Regehr, John and Chen, Yang and Cuoq, Pascal and Eide, Eric and Ellison, Chucky and Yang, Xuejun},
  booktitle={Proceedings of the 33rd ACM SIGPLAN conference on Programming Language Design and Implementation},
  pages={335--346},
  year={2012}
}

@inproceedings{perses,
  title={Perses: Syntax-guided program reduction},
  author={Sun, Chengnian and Li, Yuanbo and Zhang, Qirun and Gu, Tianxiao and Su, Zhendong},
  booktitle={Proceedings of the 40th International Conference on Software Engineering},
  pages={361--371},
  year={2018}
}

@article{yarpgen,
  title={Random testing for C and C++ compilers with YARPGen},
  author={Livinskii, Vsevolod and Babokin, Dmitry and Regehr, John},
  journal={Proceedings of the ACM on Programming Languages},
  volume={4},
  number={OOPSLA},
  pages={1--25},
  year={2020},
  publisher={ACM New York, NY, USA}
}

@article{yarpgen2.0,
  title={Fuzzing loop optimizations in compilers for C++ and data-parallel languages},
  author={Livinskii, Vsevolod and Babokin, Dmitry and Regehr, John},
  journal={Proceedings of the ACM on Programming Languages},
  volume={7},
  number={PLDI},
  pages={1826--1847},
  year={2023},
  publisher={ACM New York, NY, USA}
}

@inproceedings{grayc,
  title={GrayC: Greybox fuzzing of compilers and analysers for C},
  author={Even-Mendoza, Karine and Sharma, Arindam and Donaldson, Alastair F and Cadar, Cristian},
  booktitle={Proceedings of the 32nd ACM SIGSOFT International Symposium on Software Testing and Analysis},
  pages={1219--1231},
  year={2023}
}

@article{creal,
  title={Boosting Compiler Testing by Injecting Real-World Code},
  author={Li, Shaohua and Theodoridis, Theodoros and Su, Zhendong},
  journal={Proceedings of the ACM on Programming Languages},
  volume={8},
  number={PLDI},
  pages={223--245},
  year={2024},
  publisher={ACM New York, NY, USA}
}

@online{gitbisect,
  title = {Git Bisect},
  author = {Git},
  year = {2024},
  url = {https://git-scm.com/docs/git-bisect}
}

@article{deltadebugging,
  title={Simplifying and isolating failure-inducing input},
  author={Zeller, Andreas and Hildebrandt, Ralf},
  journal={IEEE Transactions on software engineering},
  volume={28},
  number={2},
  pages={183--200},
  year={2002},
  publisher={IEEE}
}

@online{treesitter,
  title = {Tree-sitter},
  year = {2022},
  url = {https://github.com/tree-sitter/tree-sitter}
}

@inproceedings{gumTree,
  title={Fine-grained and accurate source code differencing},
  author={Falleri, Jean-R{\'e}my and Morandat, Flor{\'e}al and Blanc, Xavier and Martinez, Matias and Monperrus, Martin},
  booktitle={Proceedings of the 29th ACM/IEEE international conference on Automated software engineering},
  pages={313--324},
  year={2014}
}

@online{gcov,
  title = {Gcov},
  year = {2020},
  url = {https://gcc.gnu.org/onlinedocs/gcc/Gcov.html}
}

@article{chen2022boosting,
  title={Boosting compiler testing via compiler optimization exploration},
  author={Chen, Junjie and Suo, Chenyao},
  journal={ACM Transactions on Software Engineering and Methodology (TOSEM)},
  volume={31},
  number={4},
  pages={1--33},
  year={2022},
  publisher={ACM New York, NY}
}

@inproceedings{chen2019history,
  title={History-guided configuration diversification for compiler test-program generation},
  author={Chen, Junjie and Wang, Guancheng and Hao, Dan and Xiong, Yingfei and Zhang, Hongyu and Zhang, Lu},
  booktitle={2019 34th IEEE/ACM International Conference on Automated Software Engineering (ASE)},
  pages={305--316},
  year={2019},
  organization={IEEE}
}

@online{git,
  title = {Git},
  year = {2024},
  url = {https://git-scm.com/}
}

@article{gonzalez1985clustering,
  title={Clustering to minimize the maximum intercluster distance},
  author={Gonzalez, Teofilo F},
  journal={Theoretical computer science},
  volume={38},
  pages={293--306},
  year={1985},
  publisher={Elsevier}
}

@inproceedings{perez2010using,
  title={Using BM25F for semantic search},
  author={P{\'e}rez-Ag{\"u}era, Jos{\'e} R and Arroyo, Javier and Greenberg, Jane and Iglesias, Joaquin Perez and Fresno, Victor},
  booktitle={Proceedings of the 3rd international semantic search workshop},
  pages={1--8},
  year={2010}
}

@inproceedings{sun2011towards,
  title={Towards more accurate retrieval of duplicate bug reports},
  author={Sun, Chengnian and Lo, David and Khoo, Siau-Cheng and Jiang, Jing},
  booktitle={2011 26th IEEE/ACM International Conference on Automated Software Engineering (ASE 2011)},
  pages={253--262},
  year={2011},
  organization={IEEE}
}

@inproceedings{sun2010discriminative,
  title={A discriminative model approach for accurate duplicate bug report retrieval},
  author={Sun, Chengnian and Lo, David and Wang, Xiaoyin and Jiang, Jing and Khoo, Siau-Cheng},
  booktitle={Proceedings of the 32nd ACM/IEEE International Conference on Software Engineering-Volume 1},
  pages={45--54},
  year={2010}
}

@inproceedings{nguyen2012duplicate,
  title={Duplicate bug report detection with a combination of information retrieval and topic modeling},
  author={Nguyen, Anh Tuan and Nguyen, Tung Thanh and Nguyen, Tien N and Lo, David and Sun, Chengnian},
  booktitle={Proceedings of the 27th IEEE/ACM International Conference on Automated Software Engineering},
  pages={70--79},
  year={2012}
}

@inproceedings{sohn2017fluccs,
  title={Fluccs: Using code and change metrics to improve fault localization},
  author={Sohn, Jeongju and Yoo, Shin},
  booktitle={Proceedings of the 26th ACM SIGSOFT International Symposium on Software Testing and Analysis},
  pages={273--283},
  year={2017}
}

@article{pelleg2004active,
  title={Active learning for anomaly and rare-category detection},
  author={Pelleg, Dan and Moore, Andrew},
  journal={Advances in neural information processing systems},
  volume={17},
  year={2004}
}

@inproceedings{vatturi2009category,
  title={Category detection using hierarchical mean shift},
  author={Vatturi, Pavan and Wong, Weng-Keen},
  booktitle={Proceedings of the 15th ACM SIGKDD international conference on Knowledge discovery and data mining},
  pages={847--856},
  year={2009}
}

@article{chen2023exploring,
  title={Exploring better black-box test case prioritization via log analysis},
  author={Chen, Zhichao and Chen, Junjie and Wang, Weijing and Zhou, Jianyi and Wang, Meng and Chen, Xiang and Zhou, Shan and Wang, Jianmin},
  journal={ACM Transactions on Software Engineering and Methodology},
  volume={32},
  number={3},
  pages={1--32},
  year={2023},
  publisher={ACM New York, NY}
}

@inproceedings{holmes2018causal,
  title={Causal distance-metric-based assistance for debugging after compiler fuzzing},
  author={Holmes, Josie and Groce, Alex},
  booktitle={2018 IEEE 29th International Symposium on Software Reliability Engineering (ISSRE)},
  pages={166--177},
  year={2018},
  organization={IEEE}
}

@article{woolson2005wilcoxon,
  title={Wilcoxon signed-rank test},
  author={Woolson, Robert F},
  journal={Encyclopedia of biostatistics},
  volume={8},
  year={2005},
  publisher={Wiley Online Library}
}

@online{icc,
  title = {Intel C++ Compiler},
  year = {2024},
  url = {https://www.intel.com/content/www/us/en/developer/tools/oneapi/dpc-compiler.html}
}

@article{sliwerski2005changes,
  title={When do changes induce fixes?},
  author={{\'S}liwerski, Jacek and Zimmermann, Thomas and Zeller, Andreas},
  journal={ACM sigsoft software engineering notes},
  volume={30},
  number={4},
  pages={1--5},
  year={2005},
  publisher={ACM New York, NY, USA}
}

@inproceedings{an2021reducing,
  title={Reducing the search space of bug inducing commits using failure coverage},
  author={An, Gabin and Yoo, Shin},
  booktitle={Proceedings of the 29th ACM Joint Meeting on European Software Engineering Conference and Symposium on the Foundations of Software Engineering},
  pages={1459--1462},
  year={2021}
}

@inproceedings{an2023fonte,
  title={Fonte: Finding bug inducing commits from failures},
  author={An, Gabin and Hong, Jingun and Kim, Naryeong and Yoo, Shin},
  booktitle={2023 IEEE/ACM 45th International Conference on Software Engineering (ICSE)},
  pages={589--601},
  year={2023},
  organization={IEEE}
}

@article{tang2024enhancing,
  title={Enhancing Bug-Inducing Commit Identification: A Fine-Grained Semantic Analysis Approach},
  author={Tang, Lingxiao and Ni, Chao and Huang, Qiao and Bao, Lingfeng},
  journal={IEEE Transactions on Software Engineering},
  year={2024},
  publisher={IEEE}
}

@article{spinellis2005version,
  title={Version control systems},
  author={Spinellis, Diomidis},
  journal={IEEE software},
  volume={22},
  number={5},
  pages={108--109},
  year={2005},
  publisher={IEEE}
}

@inproceedings{randrianaina2022benefits,
  title={On the benefits and limits of incremental build of software configurations: an exploratory study},
  author={Randrianaina, Georges Aaron and T{\"e}rnava, Xhevahire and Khelladi, Djamel Eddine and Acher, Mathieu},
  booktitle={Proceedings of the 44th International Conference on Software Engineering},
  pages={1584--1596},
  year={2022}
}

@article{zou2016automated,
  title={Automated duplicate bug report detection using multi-factor analysis},
  author={Zou, Jie and Xu, Ling and Yang, Mengning and Zhang, Xiaohong and Zeng, Jun and Hirokawa, Sachio},
  journal={IEICE TRANSACTIONS on Information and Systems},
  volume={99},
  number={7},
  pages={1762--1775},
  year={2016},
  publisher={The Institute of Electronics, Information and Communication Engineers}
}

@article{lin2016enhancements,
  title={Enhancements for duplication detection in bug reports with manifold correlation features},
  author={Lin, Meng-Jie and Yang, Cheng-Zen and Lee, Chao-Yuan and Chen, Chun-Chang},
  journal={Journal of Systems and Software},
  volume={121},
  pages={223--233},
  year={2016},
  publisher={Elsevier}
}

@inproceedings{wang2008approach,
  title={An approach to detecting duplicate bug reports using natural language and execution information},
  author={Wang, Xiaoyin and Zhang, Lu and Xie, Tao and Anvik, John and Sun, Jiasu},
  booktitle={Proceedings of the 30th international conference on Software engineering},
  pages={461--470},
  year={2008}
}

@inproceedings{lerch2013finding,
  title={Finding duplicates of your yet unwritten bug report},
  author={Lerch, Johannes and Mezini, Mira},
  booktitle={2013 17th European conference on software maintenance and reengineering},
  pages={69--78},
  year={2013},
  organization={IEEE}
}

@inproceedings{tang2023neural,
  title={Neural SZZ algorithm},
  author={Tang, Lingxiao and Bao, Lingfeng and Xia, Xin and Huang, Zhongdong},
  booktitle={2023 38th IEEE/ACM International Conference on Automated Software Engineering (ASE)},
  pages={1024--1035},
  year={2023},
  organization={IEEE}
}

@article{rosa2023comprehensive,
  title={A comprehensive evaluation of szz variants through a developer-informed oracle},
  author={Rosa, Giovanni and Pascarella, Luca and Scalabrino, Simone and Tufano, Rosalia and Bavota, Gabriele and Lanza, Michele and Oliveto, Rocco},
  journal={Journal of systems and software},
  volume={202},
  pages={111729},
  year={2023},
  publisher={Elsevier}
}

@article{rani2024refining,
  title={On Refining the SZZ Algorithm with Bug Discussion Data},
  author={Rani, Pooja and Petrulio, Fernando and Bacchelli, Alberto},
  journal={Empirical Software Engineering},
  volume={29},
  number={5},
  pages={115},
  year={2024},
  publisher={Springer}
}

@article{tang2025llm4szz,
  title={LLM4SZZ: Enhancing SZZ Algorithm with Context-Enhanced Assessment on Large Language Models},
  author={Tang, Lingxiao and Liu, Jiakun and Liu, Zhongxin and Yang, Xiaohu and Bao, Lingfeng},
  journal={arXiv preprint arXiv:2504.01404},
  year={2025}
}

@inproceedings{wen2016locus,
  title={Locus: Locating bugs from software changes},
  author={Wen, Ming and Wu, Rongxin and Cheung, Shing-Chi},
  booktitle={Proceedings of the 31st IEEE/ACM International Conference on Automated Software Engineering},
  pages={262--273},
  year={2016}
}

@inproceedings{ni2025legofuzz,
  title={LegoFuzz: Interleaving Large Language Models for Compiler Testing},
  author={Ni, Yunbo},
  booktitle={Companion Proceedings of the 2025 ACM SIGPLAN International Conference on Systems, Programming, Languages, and Applications: Software for Humanity},
  pages={37--39},
  year={2025}
}

@inproceedings{theodoridis2022finding,
  title={Finding missed optimizations through the lens of dead code elimination},
  author={Theodoridis, Theodoros and Rigger, Manuel and Su, Zhendong},
  booktitle={Proceedings of the 27th ACM International Conference on Architectural Support for Programming Languages and Operating Systems},
  pages={697--709},
  year={2022}
}

@article{theodoridis2024refined,
  title={Refined input, degraded output: The counterintuitive world of compiler behavior},
  author={Theodoridis, Theodoros and Su, Zhendong},
  journal={Proceedings of the ACM on Programming Languages},
  volume={8},
  number={PLDI},
  pages={671--691},
  year={2024},
  publisher={ACM New York, NY, USA}
}

@article{li2024boosting,
  title={Boosting compiler testing by injecting real-world code},
  author={Li, Shaohua and Theodoridis, Theodoros and Su, Zhendong},
  journal={Proceedings of the ACM on Programming Languages},
  volume={8},
  number={PLDI},
  pages={223--245},
  year={2024},
  publisher={ACM New York, NY, USA}
}

@misc{gcc_bugzilla_wrong_code_fixed,
  author       = {GNU Compiler Collection (GCC) Developers},
  title        = {GCC Bugzilla: Resolved Wrong-Code Bugs},
  year         = {2026},
  note         = {Accessed: 2026-05-16},
  url          = {https://gcc.gnu.org/bugzilla/buglist.cgi?bug_status=RESOLVED&cf_known_to_fail_type=allwords&cf_known_to_work_type=allwords&list_id=512342&product=gcc&query_format=advanced&resolution=FIXED&short_desc=wrong%20code&short_desc_type=allwordssubstr}
}

@article{yang2025using,
  title={Using a Sledgehammer to Crack a Nut? Revisiting Automated Compiler Fault Isolation},
  author={Yang, Yibiao and Li, Qingyang and Sun, Maolin and Wu, Jiangchang and Zhou, Yuming},
  journal={arXiv preprint arXiv:2512.16335},
  year={2025}
}

\end{document}